\documentclass[11pt,twoside]{amsart}

\usepackage[hmargin={3cm,3cm},vmargin={3cm,3cm},includehead]{geometry}
\usepackage{amsmath,amssymb}
\usepackage{amsfonts}
\usepackage{mathrsfs}

\usepackage{dsfont}

\usepackage{epic}
\usepackage{eepic}

\newtheorem{theorem}{Theorem}

\newcommand{\ds}{\displaystyle}
\def\EXP{\textrm{{\large e}}}
\newcommand{\bos}{\boldsymbol{a}}
\newcommand{\fer}{\boldsymbol{f}}
\newcommand{\kop}{\boldsymbol{k}}
\newcommand{\bosn}{\textrm{\scriptsize $\mathsf{N}$}}
\newcommand{\Nop}{\textrm{\scriptsize $\boldsymbol{\mathscr{N}}$}}
\newcommand{\fern}{\textrm{\scriptsize $\mathsf{M}$}}
\newcommand{\balg}{\mathcal{A}}
\newcommand{\alg}{\mathcal{C}}
\newcommand{\falg}{\mathcal{F}}
\newcommand{\Rop}{R}
\newcommand{\rop}{\mathsf{R}}
\newcommand{\Ryb}{\mathbf{R}}
\begin{document}

\title[Super-tetrahedra and super-algebras]{Super-tetrahedra and super-algebras}%

\author[S. Sergeev]{Sergey M. Sergeev}

\address{Faculty of Information Sciences and Engineering,
University of Canberra, Bruce ACT 2601 \newline \indent Department
of Theoretical Physics, Research School of Physical Sciences,
Australian National University, Canberra ACT 0200, Australia}
\email{ Sergey.Sergeev@canberra.edu.au,
sms105@rsphysse.anu.edu.au,}



\begin{abstract}
In this paper we give a detailed classification scheme for
three-dimensional quantum zero curvature representation and
tetrahedron equations. This scheme includes both even and odd
parity components, the resulting algebras of observables are
either Bose $q$-oscillators or Fermi oscillators.
Three-dimensional $R$-matrices intertwining variously oriented
tensor products of Bose and Fermi oscillators and satisfying
tetrahedron and super-tetrahedron equations are derived. The
$3d\to 2d$ compactification reproduces
$\mathscr{U}_q(\widehat{\textrm{gl}}(n|m))$ super-algebras and
their representation theory.
\end{abstract}

\maketitle

\section*{Introduction}

The $q$-oscillator solution of the quantum tetrahedron equation
was derived in \cite{BS05} as an interwiner of quantum local
Yang-Baxter equation with a specific Ansatz for auxiliary
matrices. However, a more fundamental zero curvature
representation of three-dimensional models is based on an
auxiliary linear problem \cite{MR0457963}. Namely, the
$q$-oscillator model can be viewed as the quantization
\cite{Melbourne,circular} of discrete three-wave equations and
their linear problem \cite{BogdanovKonopelchenko,DoliwaSantini}.
In the first section of this paper we discuss in details this
type's zero curvature representation and the tetrahedron equation
from the quantum mechanical point of view, and formulate a
classification problem for algebras of observables.

The $q$-oscillator algebra is the solution of the classification
problem for even algebras, Theorem \ref{Th-1} of the second
section. The linear problem provides a natural way to introduce
also odd algebras and extend the classification problem to mixed
case of even and odd algebras, corresponding classification
Theorem \ref{Th-2} is the subject of the second section as well.
The result of classification theorem is the existence of four
distinct automorphisms for even algebras $\balg(q^{\pm 1})$ and
odd algebras $\falg(q^{\pm 1})$ and eight (super-)tetrahedra for
them.

Bose and Fermi oscillators are ``evaluation representations'' of
formal algebras $\balg(q^{\pm 1})$ and $\falg(q^{\pm 1})$. These
representations are fixed in the third section. In the fourth
section we construct explicitly all corresponding quantum
intertwiners ($R$-matrices) and in the fifth section we discuss
briefly all eight tetrahedron equations.

Any solution of the tetrahedron equation produces a series of
solutions of the Yang-Baxter equation. In section six we remind
this scheme for the obtained Bose/Fermi inhomogeneous solutions of
the tetrahedron equations. The resulting quantum groups are in
general $\mathscr{U}_q(\widehat{\textrm{gl}}(n|m))$. This
statement is detailed in the last section for the illustrative
case of $\mathscr{U}_q(\widehat{\textrm{gl}}(2|1))$

\section{Zero curvature representation.}

The primary concept of integrability is an auxiliary linear
problem. The simplest form of local auxiliary linear problem in
the theory of quantum integrable systems in wholly discrete $2+1$
dimensional space-time is the pair of linear relations
\cite{Korepanov:1995}
\begin{equation}\label{lp}
\psi_\alpha'=a_j^{}\psi_\alpha^{}+b_j^{}\psi_\beta^{}\;,\quad
\psi_\beta'=c_j^{}\psi_\alpha^{}+d_j\psi_\beta^{}\;,
\end{equation}
where $a_j,b_j,c_j,d_j$ are elements of some local algebra
$\mathcal{C}_j$,
\begin{equation}
\mathcal{C}_j=\mathcal{C}[a_j,b_j,c_j,d_j]\;,
\end{equation}
and $\psi_\alpha^{}$, $\psi_\alpha'$, $\psi_\beta^{}$,
$\psi_\beta'$ are auxiliary linear elements from a formal left
module of a tensor power of local algebras.

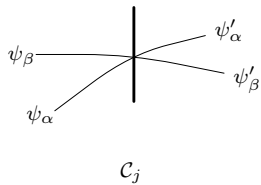
\begin{figure}[ht]
\setlength{\unitlength}{0.25mm}
\begin{picture}(140,100)
\put(20,0){\begin{picture}(100,100)
 \thinlines\spline(10,20)(50,50)(90,60)\put(-5,15){\scriptsize $\psi_\alpha$}\put(95,60){\scriptsize $\psi_\alpha'$}
 \thinlines\spline(0,50)(50,50)(100,40)\put(-15,47){\scriptsize $\psi_\beta$}\put(105,35){\scriptsize $\psi_\beta'$}
 \Thicklines\path(52,25)(52,75)\put(45,-15){\scriptsize $\mathcal{C}_j$}
\end{picture}}
\end{picture}
\caption{$3d$ geometry of auxiliary linear problem.}
\label{fig-lp}
\end{figure}

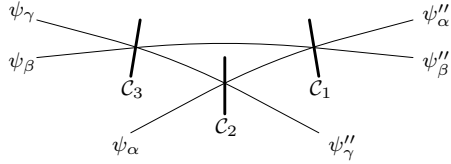
\begin{figure}[ht]
\setlength{\unitlength}{0.25mm}
\begin{picture}(240,100)
\put(20,0){\begin{picture}(200,100)
 \thinlines\spline(0,50)(100,60)(200,50)
 \thinlines\spline(0,70)(75,50)(150,10)
 \thinlines\spline(50,10)(125,50)(200,70)
 \put(-15,45){\scriptsize $\psi_\beta$}
 \put(205,45){\scriptsize $\psi_\beta''$}
 \put(-15,72){\scriptsize $\psi_\gamma$}
 \put(155,0){\scriptsize $\psi_\gamma''$}
 \put(40,0){\scriptsize $\psi_\alpha$}
 \put(205,70){\scriptsize $\psi_\alpha''$}
 \Thicklines
 \path(100,20)(100,50)\put(95,10){\scriptsize $\mathcal{C}_2$}
 \path(50,40)(55,70)\put(45,30){\scriptsize $\mathcal{C}_3$}
 \path(150,40)(145,70)\put(145,30){\scriptsize $\mathcal{C}_1$}
\end{picture}}
\end{picture}
\caption{Auxiliary linear problem for a triangle.}
\label{fig-lpleft}
\end{figure}

Geometrically, a collection of local linear problems is associated
with a $2d$ ``space-like'' section of a three-dimensional graph,
see Fig. \ref{fig-lp}. Auxiliary variables are associated with the
edges of auxiliary plane, while the elements of $\mathcal{C}_j$
are associated with the $j$th vertex of auxiliary plane which
corresponds to a ``time-like'' edge of the $3d$ graph (bold edge
in Fig. \ref{fig-lp}).
It is convenient to rewrite the local linear problem in a matrix
form,
\begin{equation}\label{X}
\left(\begin{array}{c}
\psi_\alpha'\\\psi_\beta'\end{array}\right)\;=\;
X_{\alpha\beta}[\mathcal{C}_j]\ \left(\begin{array}{c}
\psi_\alpha\\\psi_\beta\end{array}\right)\;, \qquad
X_{\alpha\beta}[\mathcal{C}_j]\;=\;\left(\begin{array}{cc} a_j &
b_j \\ c_j & d_j\end{array}\right)\;,
\end{equation}
where $X_{\alpha\beta}[\mathcal{C}]$ is a true $3d$ analogue of a
Lax operator.
Collection of local linear problems along an auxiliary $2d$ graph
relates the outer auxiliary variables. For the triangle graph in
Fig. \ref{fig-lpleft} the local linear problems allow one to
express $\psi_\alpha''$, $\psi_\beta''$ and $\psi_\gamma''$ as
linear combinations of $\psi_\alpha$, $\psi_\beta$ and
$\psi_\gamma$. These expressions can be written in the matrix form
as
\begin{equation}\label{lp-left}
\left(\begin{array}{c}
\psi_\alpha''\\\psi_\beta''\\\psi_\gamma''\end{array}\right)\;=\;
X_{\alpha\beta}[\mathcal{C}_1] X_{\alpha\gamma}[\mathcal{C}_2]
X_{\beta\gamma}[\mathcal{C}_3]\,\cdot\, \left(\begin{array}{c}
\psi_\alpha\\\psi_\beta\\\psi_\gamma\end{array}\right)\;,
\end{equation}
where $X_{\alpha\beta}$ is the two-by-two matrix (\ref{X}) in the
block $\alpha\oplus\beta$ completed by the unity in the block
$\gamma$, etc.

\begin{figure}[ht]
\setlength{\unitlength}{0.25mm}
\begin{picture}(240,100)
\put(20,0){\begin{picture}(200,100)
 \thinlines\spline(0,40)(100,50)(200,40)
 \thinlines\spline(0,20)(75,70)(150,90)
 \thinlines\spline(50,90)(125,70)(200,20)
 \put(-15,35){\scriptsize $\psi_\beta$}
 \put(205,35){\scriptsize $\psi_\beta''$}
 \put(30,90){\scriptsize $\psi_\gamma$}
 \put(205,10){\scriptsize $\psi_\gamma''$}
 \put(-10,10){\scriptsize $\psi_\alpha$}
 \put(155,90){\scriptsize $\psi_\alpha''$}
 \Thicklines
 \path(100,60)(100,90)\put(95,95){\scriptsize $\mathcal{C}_2'$}
 \path(37,30)(34,60)\put(30,65){\scriptsize $\mathcal{C}_1'$}
 \path(163,30)(166,60)\put(161,65){\scriptsize $\mathcal{C}_3'$}
\end{picture}}
\end{picture}
\caption{Opposite triangle.} \label{fig-lpright}
\end{figure}
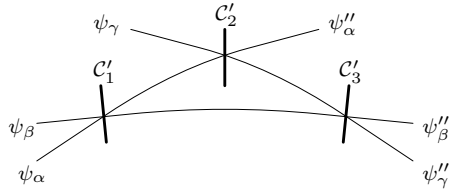

One can consider an ``opposite'' graph to Fig. \ref{fig-lpleft},
shown in Fig. \ref{fig-lpright}. The opposite graph has the same
external data as the initial one: the collection of linear
problems also allows to experess $\psi_\alpha''$, $\psi_\beta''$
and $\psi_\gamma''$ as linear combinations of $\psi_\alpha$,
$\psi_\beta$ and $\psi_\gamma$:
\begin{equation}\label{lp-right}
\left(\begin{array}{c}
\psi_\alpha''\\\psi_\beta''\\\psi_\gamma''\end{array}\right)\;=\;
X_{\beta\gamma}[\mathcal{C}_3'] X_{\alpha\gamma}[\mathcal{C}_2']
X_{\alpha\beta}[\mathcal{C}_1'] \,\cdot\, \left(\begin{array}{c}
\psi_\alpha\\\psi_\beta\\\psi_\gamma\end{array}\right)\;,
\end{equation}
with some $\mathcal{C}_j'=\mathcal{C}[a_j',b_j',c_j',d_j']$.

Zero curvature condition is the complete independence of linear
problem of a choice of graphs, Fig. \ref{fig-lpleft} and Fig.
\ref{fig-lpright}:
\begin{equation}\label{qke}
X_{\alpha\beta}[\mathcal{C}_1] X_{\alpha\gamma}[\mathcal{C}_2]
X_{\beta\gamma}[\mathcal{C}_3] \;=\;
X_{\beta\gamma}[\mathcal{C}_3'] X_{\alpha\gamma}[\mathcal{C}_2']
X_{\alpha\beta}[\mathcal{C}_1']\;.
\end{equation}
This equation resembles the local Yang-Baxter equation, however
(\ref{qke}) is the equation in tensor \emph{sum} of spaces
$\alpha,\beta$ and $\gamma$. Equation (\ref{qke}) for
$\mathbb{C}$-valued fields $\alg$ was studied in details by I.
Korepanov in \cite{Korepanov:1995}, some earlier applications of
(\ref{qke}) to functional tetrahedron equation can be found in
\cite{KashaevKorepanovSergeev}.

In classics, when $\alg_j$ are Abelian algebras, equation
(\ref{qke}) is just an equation of motion for $\mathbb{C}$-valued
fields $a_j,b_j,c_j,d_j$ since the zero curvature condition is
just a self-consistency condition for the linear problems. In the
conventional classical approach the gauge symmetry of auxiliary
fields, $\psi_\alpha\to G_\alpha\psi_\alpha$, etc., is used to
reduce the number of independent fields:
\begin{equation}
a_j=1\;,\quad b_j=-A_j^{}\;,\quad c_j=A_j^*\;,\quad
d_j=1-A_j^{}A_j^*\;,
\end{equation}
what corresponds to auxiliary relations
$\psi_\alpha^{}-\psi_\alpha'=A_j^{}\psi_\beta$ and
$\psi_\beta'-\psi_\beta^{}=A_j^*\psi_\alpha'$ of the discrete
three-wave system \cite{DoliwaSantini}.

Our aim, however, is a quantum theory where the gauge
transformations affecting quantum structure must be considered
more carefully. Yet the algebras $\mathcal{C}_j$ and
$\mathcal{C}_j'$ are uncertain. What we expect from the
foundations of quantum theories: quantum equations of motion are
conjugations by a discrete time evolution operator and therefore
the Heisenberg quantum equations of motion in discrete space-time
are sequences of automorphisms.

Following the foundations of quantum theories, suppose now that
algebras $\mathcal{C}_1$, $\mathcal{C}_2$ and $\mathcal{C}_3$ are
local and equivalent: the triplet
$[\mathcal{C}_1,\mathcal{C}_2,\mathcal{C}_3]$ is the tensor
product of three independent copies of the same algebra,
\begin{equation}\label{locality}
[\mathcal{C}_1,\mathcal{C}_2,\mathcal{C}_3]=\mathcal{C}_1\otimes
\mathcal{C}_2\otimes \mathcal{C}_3\;=\;\mathcal{C}^{\otimes 3}
\end{equation}
so that the index $j$ stands for the component of tensor
product\footnote{formally, $\mathcal{C}_j$ now stands for an
enveloping algebra of $[1,a_j,a_j^{-1},b_j,c_j,d_j,d_j^{-1}]$.}.
Then one comes to a

\noindent\textbf{Problem:} \emph{What is an algebra $\mathcal{C}$
such that equation (\ref{qke}) defines uniquely an automorphism
$\mathcal{C}^{\otimes 3}\;\to\;\mathcal{C}^{\otimes 3}$,
\begin{equation}\label{aut}
\mathcal{C}_1\otimes\mathcal{C}_2\otimes\mathcal{C}_3\ \to \
\mathcal{C}_1'\otimes\mathcal{C}_2'\otimes\mathcal{C}_3'\;.
\end{equation}
}

\noindent Natural extension of this problem is the case of
non-equivalent algebras $\alg_j$. We will prove the classification
theorems for this problem in the next section.

\begin{figure}[ht]
\begin{center}
\setlength{\unitlength}{0.25mm}
\begin{picture}(500,200)
\put(0,0)
 {\begin{picture}(200,200) \thinlines
 \drawline(90.00, 42.50)(30.00, 87.50)
 \put(20,90){\scriptsize $\gamma$}
 \drawline(62.50, 37.50)(167.50, 112.50)
 \put(170,115){\scriptsize $\alpha$}
 \drawline(12.50, 75.00)(84.44,88.08)\drawline(91.04,89.28)(177.50, 105.00)
 \put(180,100){\scriptsize $\beta$}
 \Thicklines
 \drawline(75.00, 25.00)(105.00, 175.00)
 \put(105,180){\scriptsize $2$}
 \drawline(25.00, 62.50)(115.00, 167.50)
 \put(120,170){\scriptsize $3$}
 \drawline(162.50, 87.50)(87.50, 162.50)
 \put(80,165){\scriptsize $1$}
 \end{picture}}
\put(300,0)
 {\begin{picture}(200,200) \thinlines
 \drawline(110.00, 157.50)(170.00, 112.50)
 \put(100,160){\scriptsize $\gamma$}
 \drawline(137.50, 162.50)(32.50, 87.50)
 \put(145,160){\scriptsize $\alpha$}
 \drawline(187.50, 125.00)(22.50, 95.00)
 \put(192,123){\scriptsize $\beta$}
 \Thicklines
 \drawline(125.00, 175.00)(112.85,114.25)\drawline(111.65,108.25)(95.00, 25.00)
 \put(125,185){\scriptsize $2$}
 \drawline(175.00, 137.50)(85.00, 32.50)
 \put(180,145){\scriptsize $3$}
 \drawline(37.50, 112.50)(112.50, 37.50)
 \put(30,120){\scriptsize $1$}
 \end{picture}}
\put(245,100){$=$}
\end{picture}
\end{center}
\caption{\scriptsize{Graphical representation of Equation
(\ref{intertwining}): auxiliary triangles from Figs.
\ref{fig-lpleft} and \ref{fig-lpright} with the solid vertex
standing for the intertwining operator $\Rop_{123}$. Auxiliary
planes are sections of the solid vertex of three-dimensional
lattice. Equation (\ref{intertwining}) has the structure of
tetrahedron equation in $V_1\otimes V_2\otimes V_3\otimes
(\alpha\oplus\beta\oplus\gamma)$. }} \label{fig-intertwining}
\end{figure}
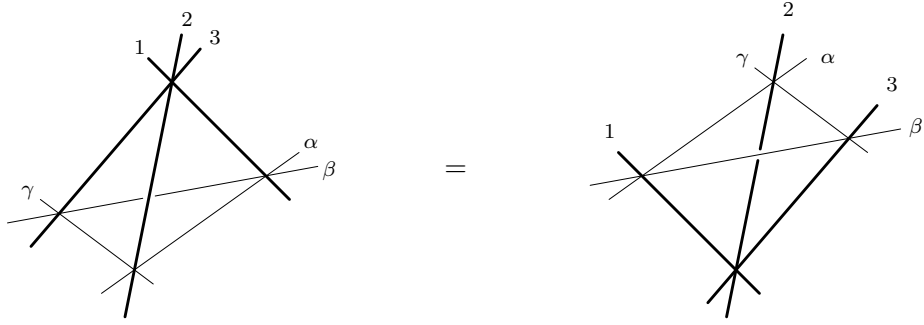

Our final aim is the explicit construction of intertwining
operators for quantum equation (\ref{qke}). If algebras $\alg_j$
and their irreducible representations are chosen properly, then
the automorphism (\ref{aut}) is internal one: there exists an
uniquely defined operator $\Rop_{123}$ such that
\begin{equation}\label{aut-conj}
\alg_j'\;=\;\Rop_{123}^{}\;\alg_j^{}\;\Rop_{123}^{-1}\;,\quad
j=1,2,3\;.
\end{equation}
Indices of $\Rop_{123}$ refer to the components of tensor product
$V_1\otimes V_2\otimes V_3$ of representation spaces of
$\alg_1\otimes\alg_1\otimes\alg_3$. Equation (\ref{qke}) takes the
form
\begin{equation}\label{intertwining}
X_{\alpha\beta}[\mathcal{C}_1] X_{\alpha\gamma}[\mathcal{C}_2]
X_{\beta\gamma}[\mathcal{C}_3]\ \Rop_{123} \;=\; \Rop_{123}\
X_{\beta\gamma}[\mathcal{C}_3] X_{\alpha\gamma}[\mathcal{C}_2]
X_{\alpha\beta}[\mathcal{C}_1]\;,
\end{equation}
see Fig. \ref{fig-intertwining} for the graphical representation
of this intertwining relation. Equation (\ref{intertwining}) can
be viewed as one of possible $3d$ extensions of Quantum group's
intertwining of co-products $R\Delta=\Delta'R$.

\begin{figure}[ht]
\begin{center}
\setlength{\unitlength}{0.25mm}
\begin{picture}(500,200)
\put(0,0)
 {\begin{picture}(200,200)
 \put(30,120){\circle*{5}}\put(70,60){\circle*{5}}\put(130,50){\circle*{5}}\put(130,150){\circle*{5}}
 \path(130,25)(130,175)\put(127,185){\scriptsize $V_1$}
 \path(52,63)(148,47)\put(150,45){\scriptsize $V_5$}
 \path(55,37.5)(145,172.5)\put(147,174.5){\scriptsize $V_3$}
 \path(5,137.5)(80,85)\path(90,78)(155,32.5)\put(157,28){\scriptsize $V_4$}
 \path(5,112.5)(155,157.5)\put(157,159.5){\scriptsize $V_2$}
 \path(18,138)(82,42)\put(10,142){\scriptsize $V_6$}
 \end{picture}}
\put(300,0)
 {\begin{picture}(200,200)
 \put(170,80){\circle*{5}}\put(130,140){\circle*{5}}\put(70,150){\circle*{5}}\put(70,50){\circle*{5}}
 \path(70,175)(70,25)\put(67,185){\scriptsize $V_1$}
 \path(52,153)(148,137)\put(150,135){\scriptsize $V_5$}
 \path(55,27.5)(112,113)\path(118,122)(145,162.5)\put(147,164.5){\scriptsize $V_3$}
 \path(45,167.5)(195,62.5)\put(197,58){\scriptsize $V_4$}
 \path(45,42.5)(195,87.5)\put(197,89.5){\scriptsize $V_2$}
 \path(182,62)(118,158)\put(110,162){\scriptsize $V_6$}
 \end{picture}}
\put(245,100){$=$}
\end{picture}
\end{center}
\caption{Graphical representation of the tetrahedron equation
(\ref{te}) in $V_1\otimes\cdots \otimes V_6$: equivalence of two
three-dimensional graphs.} \label{fig-te}
\end{figure}
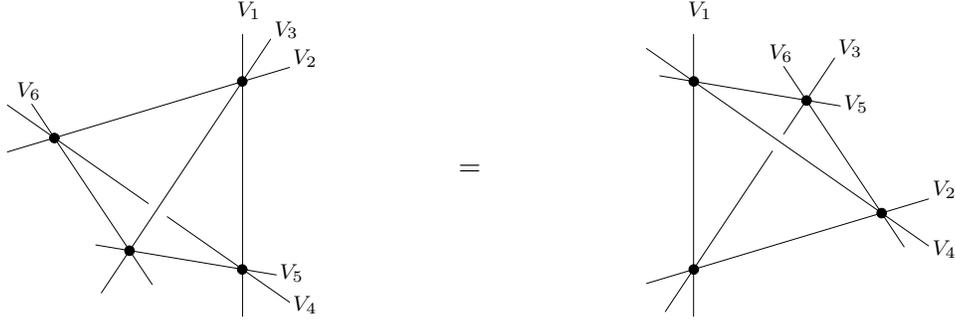

Operators $\Rop_{ijk}$ satisfy an associativity condition
following from equivalence {\small
\begin{equation}\label{quadrilateral}
X_{\alpha\beta}[\alg_1]X_{\alpha\gamma}[\alg_2]
X_{\beta\gamma}[\alg_3]X_{\alpha\delta}[\alg_4]
X_{\beta\delta}[\alg_5]X_{\gamma\beta}[\alg_6]=
X_{\gamma\beta}[\alg_6'']X_{\beta\delta}[\alg_5'']
X_{\alpha\delta}[\alg_4'']X_{\beta\gamma}[\alg_3'']
X_{\alpha\gamma}[\alg_2'']X_{\alpha\beta}[\alg_1'']
\end{equation}}
This automorphism of sixth tensor power can be decomposed into
elementary automorphisms in two different ways:
\begin{equation}
T_L=\Rop_{123}\Rop_{145}\Rop_{246}\Rop_{356}\quad
\textrm{and}\quad T_R=\Rop_{356}\Rop_{246}\Rop_{145}\Rop_{123}\;.
\end{equation}
Due to the uniqueness of automorphisms, both ways coincide:
\begin{equation}\label{te}
T_L=T_R\;,
\end{equation}
what is the tetrahedron equation -- the three-dimensional
generalization of the Yang-Baxter (triangle) equation. Graphical
representation of the tetrahedron equation is given in Fig.
\ref{fig-te}

\section{Classification theorems}

\subsection{Even case}

The answer to the problem above is
\begin{theorem}\label{Th-1}
Equation (\ref{qke}) defines an automorphism of tensor cube of
$\mathcal{C}$ is and only if
\begin{equation}\label{bose-algebra}
\mathcal{C}[a,b,c,d]=\balg(q;a,b,c,d)
\end{equation}
where $\balg(q)$ is defined by
\begin{equation}
ad=da\;,\quad ab=qba\;,\quad ca=qac\;,\quad db=qbd\;,\quad
cd=qdc\;,\quad bc-cb=(q^{-1}-q) ad\;.
\end{equation}
\end{theorem}
\noindent Algebra $\balg(q)$ has two centers,
\begin{equation}\label{balg-centers}
\eta = ad^{-1}\quad\textrm{and}\quad \xi=q^{-1}ad-bc\;.
\end{equation}
The inverse relation for $\balg(q)$ is
\begin{equation}\label{balg-inverse}
\left(\begin{array}{cc} a & b \\ c & d
\end{array}\right)^{-1}\;=\;\frac{1}{\xi} \left(\begin{array}{cc}
q^{-1}d & -b \\ -c & qa\end{array}\right)\;.
\end{equation}
We consider the centers $\eta_j$ and $\xi_j$ of $\balg_j(q)$ as
$\mathbb{C}$-numerical parameters of $\balg_j(q)$. Under this
assumption $\balg(q)$ becomes the $q$-oscillator algebra from
\cite{BS05}.

\noindent\textbf{Sketch proof.} Whichever $\mathcal{C}_j$ are
taken,
\begin{itemize}
\item the non-primed elements of
$\mathcal{C}_{j}$ with different $j$ commute since $j$ stands for
a component of tensor product, $\mathcal{C}_1=\mathcal{C}\otimes 1
\otimes 1$, etc.
\item we are looking for an automorphism. This means in
particular, primed elements of $\mathcal{C}_j'$ with different $j$
also must commute.
\end{itemize}
This is called (ultra-)locality, the starting point of the proof
is its test.

Matrix equation (\ref{qke}) in components reads:
\begin{equation}\label{qke-1}
a_2'a_1' = a_1^{} a_2^{}\;,\quad d_3' d_2' = d_2^{} d_3^{}\;,
\end{equation}
\begin{equation}\label{qke-2}
\begin{array}{ll}
\ds a_2'b_1' = b_1^{}a_3^{} + a_1^{} b_2^{} c_3^{}\;,& \ds
c_1^{}a_2^{} = a_3'c_1'+b_3'c_2'a_1'\;,\\
\ds b_2' = b_1^{}b_3^{} + a_1^{} b_2^{} d_3^{}\;,& \ds c_2^{} = c_3' c_1' + d_3' c_2' a_1' \;, \\
\ds b_3'd_2' = d_1^{} b_3^{} + c_1^{}b_2^{} d_3^{} \;, & \ds
d_2^{} c_3^{} = c_3' d_1' + d_3' c_2' b_1'
\end{array}
\end{equation}
and finally,
\begin{equation}\label{qke-3}
a_3'd_1' + b_3' c_2' b_1' = d_1^{} a_3^{} + c_1^{} b_2^{}
c_3^{}\;.
\end{equation}
Note, the left column in (\ref{qke-2}) gives almost explicit
expressions for primed $b_j$.

The locality of $\alg_1'$ and $\alg_2'$ implies in particular
\begin{equation}
[b_2',a_1^{\prime
-1}b_1']\;\equiv\;[b_1^{}b_3^{}+a_1^{}b_2^{}d_3^{},
(a_1^{}a_2^{})^{-1}(b_1^{}a_3^{}+a_1^{}b_2^{}c_3^{})]=0
\end{equation}
Expanding the commutator, we have
\begin{equation}
\begin{array}{ll}
\ds (b_1^{}a_1^{-1}\;b_3^{}a_3^{} - a_1^{-1}b_1^{}\; a_3^{}b_3^{})\; b_1^{}a_2^{-1}& \\
\ds +(b_2^{}a_2^{-1}\; d_3^{}c_3^{} - a_2^{-1}b_2^{}\; c_3^{}d_3^{})\; a_1^{}b_2^{}&\\
\ds + b_1^{}\; b_2^{}a_2^{-1} \;  d_3^{}a_3^{} -
a_1^{-1}b_1^{}a_1^{}\; a_2^{-1}b_2^{}\;  a_3^{}d_3^{} + b_1^{}\;
a_2^{-1}b_2^{}\; [b_3^{},c_3^{}] & =0
\end{array}
\end{equation}
Here the locality of $\mathcal{C}_j$ is taken into account. Check
now the structure of three lines here in $\mathcal{C}_2$. The
first line has $a_2^{-1}$, the second line has $a_2^{-1}b_2^2$,
the third line has $a_2^{-1}b_2^{}$. Since one can hardly expect
$ab\sim ba\sim a$, we can conclude that the whole expression is
zero if each line is zero. The first line gives
\begin{equation}
b_1^{}a_1^{-1}\; b_3^{}a_3^{} = a_1^{-1} b_1^{}\;
a_3^{}b_3^{}\quad \Rightarrow\quad ab=q ba
\end{equation}
for some $q$. The second line gives
\begin{equation}
b_2^{}a_2^{-1}\;d_3^{}c_3^{}=a_2^{-1}b_2^{}\;c_3^{}d_3^{}\;\quad\Rightarrow\quad
cd=q dc
\end{equation}
with the same $q$. Finally, the third line gives
\begin{equation}
[b,c]=q^{-1}ad-q da\;.
\end{equation}
In a similar way one can check
\begin{equation}
[b_2',d_3^{\prime -1}b_3']=0
\end{equation}
and get
\begin{equation}
db=q' bd\;,\quad ca=q' ac\;,\quad [b,c]=q^{\prime -1} da - q' ad
\end{equation}
for some $q'$. Comparing two variants of $[b,c]$, we have
\begin{equation}
q da = q' ad\;.
\end{equation}
All the other locality tests provide no additional information.
Thus, the locality test gives us the most general form of $\alg$:
\begin{equation}\label{Cxixi}
\alg(q,q')\;:\quad\left\{\begin{array}{l} \ds ab=q ba\;,\quad cd=q
dc\;,\quad
\ds db=q' bd\;,\quad ca = q' ac\;,\\ \\
\ds [b,c]=(q^{-1}-q') ad\;,\quad q da = q' ad
\end{array}\right.
\end{equation}
One can verify further, for $\alg_j=\alg(q,q')$ the system of
relations (\ref{qke}) is self-consistent: not only locality but
relations (\ref{Cxixi}) for primed elements do not contradict the
system (\ref{qke}). However, we have no uniqueness yet. Nine
equations (\ref{qke}) for twelve elements do not produce a unique
map in general. The map is unique only if $\alg(q,q')$ has centers
and the map conserves them (centers of algebras commute with their
intertwiner, Eq. (\ref{aut-conj})) . It is easy to verify the
following

\noindent\textbf{Lemma:} \emph{The algebra $\alg(q,q')$ has
centers compatible with (\ref{qke}) if and only if $q=q'$.}
\noindent This lemma can be proven in quasiclassical limit
$q=\EXP^{\hslash}\to 1$ and $q'=\EXP^{\hslash'}\to 1$,  so that
$\hslash/\hslash'=\delta$ as a parameter of Poisson algebra.
Equation for a center of Poisson algebra is a system of
differential equation with trivial solution unless $\delta\neq \pm
1$. Conservation of centers for the case $\delta=-1$, what is
$q'=q^{-1}$ and $[b,c]=0$, contradict with (\ref{qke}). Thus
$q=q'$ is the only choice.

Final step: when the centers (\ref{balg-centers}) conserve,
\begin{equation}
\textrm{centers of } \balg_j = \textrm{centers of } \balg_j'\quad
\textrm{for all } j\;,
\end{equation}
we can solve (\ref{qke-1},\ref{qke-2},\ref{qke-3}) with respect to
all primed generators (equation (\ref{balg-inverse}) is of
exceptional use) and verify directly that this is the automorphism
of $\balg(q)^{\otimes 3}$.  \hfill $\blacksquare$

Remark: in this proof we initially considered $\alg^{\otimes 3}$
framework -- the tensor cube of the same algebra. However, the
answer is the same without this assumption; the equivalence of
$\alg_j$ follows from analysis of all possible locality relations.

Another remark concerns the case $q'\neq q$. In general,
$\alg(q,q')=\textrm{Qsc}_{qq'}\otimes\textrm{Weyl}_{q/q'}$, the
Weyl subalgebra is generated by $\eta$ and $\xi$,
$\eta\,\cdot\,\xi=\left(\frac{q}{q'}\right)^2\xi\,\cdot\,\eta$.
For instance, the algebras $\alg(q^2,1)$ and $\alg(1,q^2)$ appear
in the three-dimensional approach to spectral equations
\cite{Sergeev:2005d}. Due to ambiguity, equation
(\ref{intertwining}) does not define intertwiner $R$ for
$\alg(q,q')^{\otimes 3}$ uniquely. However, such intertwiner
satisfying the tetrahedron equation can be constructed by dressing
the constant $q$-oscillator $R$-matrix (Eq. (\ref{raaa}) below) by
non-commutative exponential factors.

\subsection{Odd case}

Theorem \ref{Th-1} is based on the locality principle: elements in
different components of a tensor product commute. This is the
feature of even algebras; for odd algebras their odd elements in
different components of a tensor product anti-commute.

A natural (and only possible) way to introduce odd algebras is to
assign a parity to auxiliary variables, for instance to modify the
linear problem as follows:
\begin{equation}\label{lp-2}
\psi_\alpha'=a_j^{}\psi_\alpha^{}+\underline{b_j}^{}\psi_{\underline{\beta}}^{}\;,\quad
\psi_{\underline{\beta}}'=\underline{c_j}^{}\psi_\alpha^{}+d_j\psi_{\underline{\beta}}^{}\;.
\end{equation}
Here the under-line symbol marks the odd parity, relations
(\ref{lp-2}) correspond to the change of parity of auxiliary
$\beta$-line. Thus, the odd algebras correspond to parity changes
of some of the lines $\alpha$, $\beta$ or $\gamma$ in equation
(\ref{qke}); in general there are eight different parity patterns
of $\alpha,\beta,\gamma$. This way is the only possible one since
it takes into account the parity conservation in equation
(\ref{qke}). Now we are ready to extend Theorem \ref{Th-1}:

\begin{theorem}\label{Th-2} Given $X_{\alpha\beta}=X_{\alpha\beta}[\balg(q)]$, equations (\ref{qke})
provide automorphisms if and only if
\begin{equation}\label{patterns}
X_{\alpha\underline{\beta}}=X_{\alpha\underline{\beta}}[\falg(q)]\;,\quad
X_{\underline{\alpha}\beta}=X_{\underline{\alpha}\beta}[\falg(q^{-1})]\;,\quad
X_{\underline{\alpha\beta}}=X_{\underline{\alpha\beta}}[\balg(q^{-1})]
\end{equation}
for all parity patterns, where
$\falg(q)=\falg(q;a,\underline{b},\underline{c},d)$ (odd elements
are underlined) is defined by
\begin{equation}\label{Falg-def}
ad=da\;,\quad a\underline{b}=q\underline{b}a\;,\quad
\underline{c}a=qa\underline{c}\;,\quad
\underline{b}d=qd\underline{b}\;,\quad
d\underline{c}=q\underline{c}d\;,\quad
\underline{b}\underline{c}+\underline{c}\underline{b}=(q-q^{-1})
ad\;.
\end{equation}
\end{theorem}

\bigskip

\noindent For the odd elements of $\falg(q^{\pm 1})$ we have
\begin{equation}
\underline{b}^2=\underline{c}^2=0\;.
\end{equation}
Algebra $\falg(q)$ has two centers $\xi$ and $\eta$ as well,
\begin{equation}\label{f-c1}
\xi = ad\;,
\end{equation}
and $\eta$ is defined by
\begin{equation}\label{f-c2}
qad-\underline{b}\underline{c}=\eta d^2\;,\quad
qad-\underline{c}\underline{b} = \eta^{-1}a^2\;.
\end{equation}
The inversion relation for $\falg(q)$ is
\begin{equation}\label{f-inverse}
\left(\begin{array}{cc} a & \underline{b} \\ \underline{c} & d
\end{array}\right)^{-1}\;=\;\frac{q}{\xi} \left(\begin{array}{cc}
\eta^{-1}a & -\underline{b} \\ -\underline{c} & \eta
d\end{array}\right)\;.
\end{equation}
We consider the centers $\eta_j$ and $\xi_j$ of $\falg_j(q)$ as
$\mathbb{C}$-numerical parameters of $\falg_j(q)$. Note also the
existence of two orthogonal projectors in $\falg(q)$:
\begin{equation}
P_1=qa-\eta d\;,\quad P_2=qd-\eta^{-1}a\;,
\end{equation}
such that
\begin{equation}
P_1P_2= \underline{b}P_1 =
P_1\underline{c}=\underline{c}P_2=P_2\underline{b}=0\;.
\end{equation}

\noindent{\textbf{Sketch proof of Theorem \ref{Th-2}.}} Consider
as an example the parity pattern $\alpha\underline{\beta}\gamma$.
By the condition of the theorem, $\mathcal{C}_2=\balg_2(q)$ while
$\mathcal{C}_1$ and $\mathcal{C}_3$ are uncertain, but  $c_{1,3}$
and $b_{1,3}$ are odd elements.

The locality test: whichever $\mathcal{C}_1$ and $\mathcal{C}_3$
are taken, the ultra-locality demands
\begin{equation}\label{ultra-loc}
\left( a_2\underline{b}_1 b_2 - q b_2 a_2 \underline{b}_1
\right)'=\left(\underline{b}_3d_2b_2-qb_2\underline{b}_3d_2\right)'=\left(
a_2\underline{b}_1\underline{b}_3d_2 +
\underline{b}_3d_2a_2\underline{b}_1\right)'=0\;.
\end{equation}
Here we take into account $\mathcal{C}_2'=\balg_2'(q)$ and the
parity of $\mathcal{C}_1'$ and $\mathcal{C}_3'$. Immediate
consequence of (\ref{ultra-loc}) is
\begin{equation}\label{fermi-1-form}
a_1\underline{b}_1=q\underline{b}_1a_1\;,\quad \underline{c}_1 a_1
= q a_1 \underline{c}_1\;,\quad
\underline{c}_1\underline{b}_1+\underline{b}_1\underline{c}_1=(q-q^{-1})
a_1d_1
\end{equation}
and
\begin{equation}\label{fermi-3-form}
\underline{c}_3d_3=qd_3\underline{c}_3\;,\quad d_3 \underline{b}_3
= q \underline{b}_3 d_3\;,\quad
\underline{c}_3\underline{b}_3+\underline{b}_3\underline{c}_3=(q^{-1}-q)
a_3d_3\;.
\end{equation}
Therefore, $\mathcal{C}_1=\falg_1(q^{-1})$ and
$\mathcal{C}_3=\falg_3(q)$ according to the definition
(\ref{Falg-def}).

The final step of the proof: when the centers (\ref{f-c1}) and
(\ref{f-c2}) conserve,
\begin{equation}
\textrm{centers of }\alg_j = \textrm{centers of }\alg_j'\quad
\textrm{for all } j
\end{equation}
then equations (\ref{qke}) can be solved with respect to all
primed elements (again with an intensive use of inversion relation
(\ref{f-inverse})) and one can verify directly that (\ref{qke})
provide an unique automorphism of
\begin{equation}
\underbrace{\falg_1(q)}_{\alpha\underline{\beta}}\otimes
\underbrace{\balg_2(q)}_{\alpha\gamma}\otimes
\underbrace{\falg_3(q^{-1})}_{\underline{\beta}\gamma}\;.
\end{equation}
All the other parity patterns can be considered similarly. \hfill
$\blacksquare$

\section{``Evaluation representations'' of $\balg(q^{\pm 1})$ and $\falg(q^{\pm
1})$}

Before the derivation of intertwining operators we must choose
firstly irreducible representations of $\balg(q^{\pm 1})$ and
$\falg(q^{\pm 1})$. Algebra $\balg$ is equivalent to Bose
$q$-oscillator extended by two $\mathbb{C}$-valued parameters
(centers of $\balg(q)$). Algebra $\falg$ is equivalent to Fermi
oscillator with two extra $\mathbb{C}$-valued parameters. In all
considerations below we imply $0<|q|<1$.

\subsection{Bose $q$-oscillator}

We define the Bose oscillator by
\begin{equation}
\bos^+\bos^-=1-q^{2\bosn}\;,\quad
\bos^-\bos^+=1-q^{2\bosn+2}\;,\quad [\bosn,\bos^{\pm}]=\pm
\bos^{\pm}\;.
\end{equation}
For shortness we use
\begin{equation}
q^\bosn=\kop\;.
\end{equation}
The Fock space representations are defined either by
\begin{equation}\label{fock}
\bos^-|0\rangle \;=\;0\;,\quad
\textrm{Sectrum}(\bosn)=0,1,2,3,\dots
\end{equation}
where $|0\rangle$ is the Fock vacuum, or by
\begin{equation}\label{anti-fock}
\bos^+|-1\rangle \;=\;0\;,\quad
\textrm{Sectrum}(\bosn)=-1,-2,-3,-4,\dots
\end{equation}
where $|-1\rangle$ is an ``anti-vacuum''. These two
representations are related by the external automorphism $\iota$,
\begin{equation}\label{iota}
\iota(\kop)=q^{-1}\kop^{-1}\;,\quad
\iota(\bos^+)=\bos^-\kop^{-1}\;,\quad
\iota(\bos^-)=-\kop^{-1}\bos^+\;.
\end{equation}
Here we prefer not to fix scales of $\bos^{\pm}$ and use
$\bos^{\pm}\to \xi^{\pm 1} \bos^{\pm}$ invariant formalism.

\subsection{Fermi oscillator}

Fermi oscillator is defined by
\begin{equation}
[\fer^+,\fer^-]_+ = (1-q^2) \;,\quad
[\fern,\fer^{\pm}]=\pm\fer^{\pm}\;,\quad
(\fer^{+})^{2}=(\fer^{-})^{2}=0\;,
\end{equation}
where $[\;,\;]_+$ stands for anti-commutator. Fields $\fer^{\pm}$
have odd parity. For instance, the locality of two copies
$\falg_1$ and $\falg_2$ of Fermi oscillators means
\begin{equation}
\fer_1\fer_2+\fer_2\fer_1=0\;,\quad \textrm{etc.}
\end{equation}
The Fock vacuum is annihilated by $\fer^-$. Since
$\textrm{Spectrum}(\fern)=0\;\textrm{and}\;1$, the Fock space
representation of Fermi oscillator implies in addition
$g(\fern)=g(0)(1-\fern)+g(1)\fern$,
\begin{equation}
\fern^2=\fern\;,\quad \fern\fer^-=\fer^+\fern=0\;,\quad
\fer^-\fern=\fer^-\;,\quad \fern\fer^+=\fer^+\;,
\end{equation}
and
\begin{equation}
\fer^+\fer^-=(1-q^2)\fern\;,\quad \fer^-\fer^+=(1-q^2)(1-\fern)\;.
\end{equation}
Let for the shortness again
\begin{equation}
\kop=q^{\fern}=1-(1-q)\fern\;.
\end{equation}
Automorphism (\ref{iota}) for Fermi oscillator,
$\iota(\fern)=1-\fern$ and  $\iota(\fer^{\pm})=\fer^{\mp}$, is the
internal one and therefore there is no necessity to consider it
separately.

\subsection{Representation of $\balg(q)$}

We choose the following form of matrix $X[\balg(q)]$:
\begin{equation}\label{xbalg1}
X[\balg(q)]\;=\;\left(\begin{array}{cc} \lambda \kop & \bos^+ \\\\
\lambda\mu \bos^- & -q\mu\kop
\end{array}\right)\;,\quad X[\balg(q)]^{-1}=\left(\begin{array}{cc} \kop/{\lambda} &
\bos^+/{\lambda\mu}\\\\
\bos^- &  -q\kop/\mu\end{array}\right)\;,
\end{equation}
where $q$-oscillator is either in (\ref{fock}) or in
(\ref{anti-fock}) representation. In this parameterization
$\eta=-\frac{\lambda}{q\mu}$ and $\xi=-\lambda\mu$.

\subsection{Representation of $\balg(q^{-1})$}

We choose
\begin{equation}\label{xbalg2}
X[\balg(q^{-1})]\;=\;\left(\begin{array}{cc} q\lambda \kop &
\bos^-
\\\\ \lambda\mu \bos^+ & -\mu\kop
\end{array}\right)\;,\quad X[\balg(q^{-1})]^{-1}=\left(\begin{array}{cc} q\kop/{\lambda} &
\bos^-/{\lambda\mu}\\\\
\bos^+ &  -\kop/\mu\end{array}\right)\;.
\end{equation}

\subsection{Representation of $\falg(q)$}

We choose
\begin{equation}\label{xfalg1}
X[\falg(q)]\;=\;\left(\begin{array}{cc} \lambda\kop & \fer^+
\\\\
\lambda\mu \fer^- & -q\mu\kop^{-1}\end{array}\right)\;,\quad
X[\falg(q)]^{-1}\;=\;\left(\begin{array}{cc}
\kop/{\lambda} & \fer^+/{\lambda\mu} \\
\\ \fer^- &
-q\kop^{-1}/\mu\end{array}\right)\;.
\end{equation}
In this parameterization $\eta=-\frac{\lambda}{\mu}$ and
$\xi=-q\lambda\mu$.

\subsection{Representation of $\falg(q^{-1})$}

We choose
\begin{equation}\label{xfalg2}
X[\falg(q^{-1})]\;=\;\left(\begin{array}{cc} q^{-1}\lambda\kop &
q^{-1}\fer^-
\\\\
-q^{-1}\lambda\mu \fer^+ & -\mu\kop^{-1}\end{array}\right)\;,\quad
X[\falg(q^{-1})]^{-1}\;=\;\left(\begin{array}{cc}
q^{-1}\kop/\lambda & -q^{-1}\fer^-/{\lambda\mu} \\
\\ -q^{-1} \fer^+ &
-\kop^{-1}/\mu\end{array}\right)\;.
\end{equation}

\subsection{Remarks.}

Linear equation (\ref{X}) for matrix $X[\balg(q)]$ (\ref{xbalg1})
may be rewritten identically as
\begin{equation}
\left(\begin{array}{c}
\psi_\beta'\\\psi_\alpha^{}\end{array}\right)\;=\;
\widetilde{X}_{\beta\alpha}\ \left(\begin{array}{c}
\psi_\beta^{}\\\psi_\alpha'\end{array}\right)\;, \qquad
\widetilde{X}_{\beta\alpha}\;=\;\left(\begin{array}{cc} -\mu
q^{-1}\kop^{-1} & \mu\bos^-\kop^{-1}
\\ -\lambda^{-1}\kop^{-1}\bos^+ & \lambda^{-1}\kop^{-1}\end{array}\right)\;,
\end{equation}
so that $\widetilde{X}_{\beta\alpha}\;\sim\;\iota(
X_{\alpha\beta})$ up to a change of spectral parameters. Thus, a
change of orientation of linear problem (see Fig. \ref{fig-lp}) is
equivalent to $\iota$-automorphisms (\ref{iota}).

\section{Eight intertwining relations}

In this section we give the list of intertwining operators
(\ref{intertwining}) for all parity patterns of
$\alpha,\beta,\gamma$ for representations of $\balg(q^{\pm 1})$,
$\falg(q^{\pm 1})$ chosen above. For shortness, the symbols
$\balg$ and $\falg$ as the indices of three-dimensional
$R$-matrices will stand for corresponding representation Fock
spaces.

The following eight intertwining relations hold.

\subsection{Configuration $\alpha\beta\gamma$}
The basic intertwining relation is
\begin{equation}\label{xxxr}
X_{\alpha\beta}[\balg_1^{}(q)] X_{\alpha\gamma}[\balg_2^{}(q)]
X_{\beta\gamma}[\balg_3^{}(q)] \ \Rop = \Rop\
X_{\beta\gamma}[\balg_3(q)] X_{\alpha\gamma}[\balg_2(q)]
X_{\alpha\beta}[\balg_1(q)]\;.
\end{equation}
The intertwiner here is
\begin{equation}\label{Raaa-1}
\Rop\;=\;\Rop_{\balg_1(q)\balg_2(q)\balg_3(q)}(u,v,w)\;=\;
v^{\bosn_2} \rop_{\balg_1\balg_2\balg_3}^{} u^{\bosn_1}
w^{\bosn_3}\;,
\end{equation}
where for brevity
\begin{equation}\label{uvw}
u\;=\;\frac{\lambda_3}{\lambda_2}\;,\quad
v\;=\;\lambda_1\mu_3\;,\quad w\;=\;\frac{\mu_1}{\mu_2}\;.
\end{equation}
This definition of parameters $u,v,w$ is used in all cases below.
Constant matrix $\rop$ can be written as a power series
\begin{equation}\label{raaa}
\rop_{\balg_1\balg_2\balg_3}=R_0(\bosn_1,\bosn_2,\bosn_3)+\sum_{k=1}^{\infty}
\left(R_k(\bosn_1,\bosn_2,\bosn_3)(\bos_1^{-}\bos_2^+\bos_3^{-})^k
+ (\bos_1^+\bos_2^{-}\bos_3^+)^k
R_k(\bosn_1,\bosn_2,\bosn_3)\right)\;,
\end{equation}
where the coefficients $R_k(\bosn_1,\bosn_2,\bosn_3)$ are given by
expansion of three equivalent generating functions:
\begin{equation}\label{tr1}
\mathop{\textrm{Tr}}_{\balg_1} \left(z^{\bosn_1} \bos_1^{- k}
\bos_1^{+ k} R_k(\bosn_1,\bosn_2,\bosn_3)\right)\;=\;
z^{\bosn_2-k} \frac{ (q^{\bosn_3-\bosn_2+k+2} z^{-1};q^2)_\infty}
{(q^{\bosn_3+\bosn_2-k+2} z^{-1};q^2)_\infty}
\frac{(q^{\bosn_3+\bosn_2+k+2} z;q^2)_\infty}
{(q^{\bosn_3-\bosn_2+k} z;q^2)_\infty}\;,
\end{equation}
\begin{equation}\label{tr2}
\mathop{\textrm{Tr}}_{\balg_2} \left(z^{\bosn_2} \bos_2^{- k}
R_k(\bosn_1,\bosn_2,\bosn_3) \bos_2^{+ k}\right) \;=\;
q^{\bosn_1\bosn_3}\frac{ (-q^{1-\bosn_1-\bosn_3}z;q^2)_\infty
(-q^{3+\bosn_1+\bosn_3+2k}z;q^2)_\infty}
{(-q^{1-\bosn_1+\bosn_3}z;q^2)_\infty
(-q^{1+\bosn_1-\bosn_3}z;q^2)_\infty}\;,
\end{equation}
and
\begin{equation}\label{tr3}
\mathop{\textrm{Tr}}_{\balg_3} \left(z^{\bosn_3} \bos_3^{- k}
\bos_3^{+ k} R_k(\bosn_1,\bosn_2,\bosn_3)\right) \;=\;
z^{\bosn_2-k} \frac{ (q^{\bosn_1-\bosn_2+k+2} z^{-1};q^2)_\infty}
{(q^{\bosn_1+\bosn_2-k+2} z^{-1};q^2)_\infty}
\frac{(q^{\bosn_1+\bosn_2+k+2} z;q^2)_\infty}
{(q^{\bosn_1-\bosn_2+k} z;q^2)_\infty}\;.
\end{equation}
Here we use the standard $q$-hypergeometric notations \cite{GR}:
\begin{equation}
(x;q^2)_n=\prod_{k=0}^{n-1} (1-xq^{2k})\;,\quad (x;q^2)_\infty =
\prod_{k=0}^\infty (1-xq^{2k})\;.
\end{equation}
These generating functions are valid for both representations
(\ref{fock}) and (\ref{anti-fock}) for any of
$\balg_1,\balg_2,\balg_3$. Note, for finite integer $\bosn_j$ and
$k$ all generating functions are rational functions regular at
$z=0$ and at $z=\infty$ with finite number of poles in $z$-plane,
a generation function $F(z)=\mathop{\textrm{Tr}}_{\balg} z^{\bosn}
f(\bosn)$ gives $f(\bosn)$ as expansion near $z=0$ or near
$z=\infty$ for representations (\ref{fock}) or (\ref{anti-fock})
respectively. The expressions for $R_k(\bosn_1,\bosn_2,\bosn_3)$
in terms of $q$-hypergeometric function can be obtained by Cauchy
integrals of generating functions, in particular the matrix
elements of (\ref{raaa}) for representation (\ref{fock}) in all
$\balg_1,\balg_2,\balg_3$ are given in \cite{BS05,circular}.

Note a few symmetry properties. Firstly, generating functions
provide the definition of $R_k$ for negative $k$,
\begin{equation}
R_{-k}(n_1+k,n_2-k,n_3+k) \;=\;
\frac{(q^2;q^2)_{n_1+k}(q^2;q^2)_{n_2}(q^2;q^2)_{n_3+k}}{(q^2;q^2)_{n_1}(q^2;q^2)_{n_2-k}(q^2;q^2)_{n_3}}\;
R_k(n_1,n_2,n_3)\;.
\end{equation}
Power series (\ref{raaa}) can be formally rewritten as
\begin{equation}
\rop_{\balg_1\balg_2\balg_3}=\sum_{k=-\infty}^\infty
R_k(\bosn_1,\bosn_2,\bosn_3)(\bos_1^{-}\bos_2^+\bos_3^{-})^k =
\sum_{k=-\infty}^\infty (\bos_1^+\bos_2^{-}\bos_3^+)^k
R_k(\bosn_1,\bosn_2,\bosn_3)\;,
\end{equation}
where $R_k\equiv 0$ at states where negative powers of creation
and annihilation operators are not defined.

Operator (\ref{raaa}) is the square root of unity,
\begin{equation}
\rop_{\balg_1\balg_2\balg_3}^{-1}=\rop_{\balg_1\balg_2\balg_3}^{}
\end{equation}
for any choice of representations (\ref{fock},\ref{anti-fock}).
Remarkably, an analytic proof of this statement involves the
Ramanujan summation formula \cite{GR}. Also, expression
(\ref{raaa}) has the evident symmetry with respect to an
anti-involution $\bosn\to\bosn$ and $\bos^{\pm}\to\bos^{\mp}$.
This anti-involution is the Hermitian conjugation for real $q$ and
unitary representation (\ref{fock}).

Operator (\ref{raaa}) is the unique solution of (\ref{xxxr})
provided the integer spectra of $\bosn_j$.

\subsection{Configuration $\underline{\alpha\beta\gamma}$}

Relation {\small
\begin{equation}
X_{\underline{\alpha\beta}}[\balg_1(q^{-1})]
X_{\underline{\alpha\gamma}}[\balg_2(q^{-1})]
X_{\underline{\beta\gamma}}[\balg_3(q^{-1})] \ \Rop = \Rop \
X_{\underline{\beta\gamma}}[\balg_3(q^{-1})]
X_{\underline{\alpha\gamma}}[\balg_2(q^{-1})]
X_{\underline{\alpha\beta}}[\balg_1(q^{-1})]
\end{equation}}
provides
\begin{equation}\label{Raaa-2}
\Rop\;=\;\Rop_{\balg_1(q^{-1})\balg_2(q^{-1})\balg_3(q^{-1})}(u,v,w)
\;=\;v^{-\bosn_2} \rop_{\balg_1\balg_2\balg_3}^{} u^{-\bosn_1}
w^{-\bosn_3}\;.
\end{equation}
where $u,v,w$ and $\rop_{\balg_1\balg_2\balg_3}^{}$ are given by
(\ref{uvw}) and (\ref{raaa}).

\subsection{Fermionic configurations and intertwiners}

Intertwining operators for the fermions can be presented in two
ways. One is the way of matrix elements for the basis of fermionic
states, even vacuum $|0\rangle$ and odd one-fermion state
$|1\rangle = \frac{\fer^+}{\sqrt{1-q^2}} |0\rangle$; all equations
in matrix elements will have then unpleasant sign factors taking
into account the odd parity of the state $|1\rangle$ and ordering
of fermions. Another way avoiding the sign factors is to consider
fermionic operators directly as expressions in terms of odd
fermionic creation and annihilation operators like in
\cite{Shiroishi:1998F}. We choose the second way here. Besides,
the expressions for the fermionic intertwiners below do not need
integer spectra of Bose oscillators and therefore they are valid
also for the modular representation of $q$-oscillator
\cite{circular}.

\subsection{Configuration $\underline{\alpha}\beta\gamma$}

Relation
\begin{equation}
X_{\underline{\alpha}\beta}[\falg_1(q^{-1})]
X_{\underline{\alpha}\gamma}[\falg_2(q^{-1})]
X_{\beta\gamma}[\balg_3(q)] \ \Rop = \Rop \
X_{\beta\gamma}[\balg_3(q)]
X_{\underline{\alpha}\gamma}[\falg_2(q^{-1})]
X_{\underline{\alpha}\beta}[\falg_1(q^{-1})]
\end{equation}
yields uniquely
\begin{equation}\label{Rffa-1}
\Rop\;=\;\Rop_{\falg_1(q^{-1})\falg_2(q^{-1})\balg_3(q)}(u,v,w) =
v^{-\fern_2}\;\rop_{\falg_1\falg_2\balg_3}^{}\;
u^{-\fern_1}w^{\bosn_3}
\end{equation}
where the constant matrix $\rop$ is
\begin{equation}\label{rffa}
\rop_{\falg_1\falg_2\balg_3}^{} = (1-\fern_1)(1-\fern_2) +
q\fern_1(1-\fern_2)\kop_3 - (1-\fern_1)\fern_2\kop_3 -
\fern_1\fern_2 +
\frac{\fer_1^+\fer_2^{-}\bos_3^{-}-\fer_1^{-}\fer_2^{+}\bos_3^{+}}{1-q^2}
\end{equation}
and $u,v,w$ are given by (\ref{uvw}). Constant $R$-matrix is the
root of unity,
\begin{equation}
\rop_{\falg_1\falg_2\balg_3}^{}\;=\;
\rop_{\falg_1\falg_2\balg_3}^{-1}\;,
\end{equation}
and expression (\ref{rffa}) is symmetric with respect to
$\fer^{\pm}\to\fer^{\mp}$, $\bos^{\pm}\to\bos^{\mp}$
anti-involution.

\subsection{Configuration $\alpha\underline{\beta\gamma}$}

Relation
\begin{equation}
X_{\alpha\underline{\beta}}[\falg_1(q)]
X_{\alpha\underline{\gamma}}[\falg_2(q)]
X_{\underline{\beta\gamma}}[\balg_3(q^{-1})] \ \Rop = \Rop \
X_{\underline{\beta\gamma}}[\balg_3(q^{-1})]
X_{\alpha\underline{\gamma}}[\falg_2(q)]
X_{\alpha\underline{\beta}}[\falg_1(q)]
\end{equation}
yields
\begin{equation}\label{Rffa-2}
\Rop\;=\;\Rop_{\falg_1(q)\falg_2(q)\balg_3(q^{-1})}^{}(u,v,w) =
v^{\fern_2} \; \rop_{\falg_1\falg_2\balg_3}\; u^{\fern_1}
w^{-\bosn_3}
\end{equation}
where the constant matrix $\rop_{\falg_1\falg_2\balg_3}$ is given
by (\ref{rffa}).

\subsection{Configuration $\alpha\beta\underline{\gamma}$}

Relation
\begin{equation}
X_{\alpha\beta}[\balg_1(q)]
X_{\alpha\underline{\gamma}}[\falg_2(q)]
X_{\beta\underline{\gamma}}[\falg_3(q)] \ \Rop = \Rop \
X_{\beta\underline{\gamma}}[\falg_3(q)]
X_{\alpha\underline{\gamma}}[\falg_2(q)]
X_{\alpha\beta}[\balg_1(q)]
\end{equation}
provides uniquely
\begin{equation}\label{Raff-1}
\Rop\;=\;\Rop_{\balg_1(q)\falg_2(q)\falg_3(q)}(u,v,w)\;=\;
v^{\fern_2} \rop_{\balg_1\falg_2\falg_3}^{} u^{\bosn_1}
w^{\fern_3}
\end{equation}
where
\begin{equation}\label{raff}
\rop_{\balg_1\falg_2\falg_3}^{}=(1-\fern_2)(1-\fern_3) -
q\kop_1\fern_2(1-\fern_3) + \kop_1(1-\fern_2)\fern_3 -
\fern_2\fern_3 +
\frac{\bos_1^{-}\fer_2^+\fer_3^{-}-\bos_1^+\fer_2^{-}\fer_3^{+}}{1-q^2}
\end{equation}
Operator (\ref{raff}) is the square root of unity symmetric with
respect to the anti-involution.

\subsection{Configuration $\underline{\alpha\beta}\gamma$}

Relation {\small
\begin{equation}
X_{\underline{\alpha\beta}}[\balg_1(q^{-1})]
X_{\underline{\alpha}\gamma}[\falg_2(q^{-1})]
X_{\underline{\beta}\gamma}[\falg_3(q^{-1})] \ \Rop = \Rop \
X_{\underline{\beta}\gamma}[\falg_3(q^{-1})]
X_{\underline{\alpha}\gamma}[\falg_2(q^{-1})]
X_{\underline{\alpha\beta}}[\balg_1(q^{-1})]
\end{equation}}
gives
\begin{equation}\label{Raff-2}
\Rop\;=\;\Rop_{\balg_1(q^{-1})\falg_2(q^{-1})\falg_3(q^{-1})}^{}(u,v,w)\;=\;
v^{-\fern_2} \rop_{\balg_1\falg_2\falg_3}^{} u^{-\bosn_1}
w^{-\fern_3}
\end{equation}
where $\rop_{\balg_1\falg_2\falg_3}^{}$ is given by (\ref{raff}).

\subsection{Configuration $\alpha\underline{\beta}\gamma$}

Relation
\begin{equation}
X_{\alpha\underline{\beta}}[\falg_1(q)]
X_{\alpha\gamma}[\balg_2(q)]
X_{\underline{\beta}\gamma}[\falg_3(q^{-1})] \ \Rop = \Rop \
X_{\underline{\beta}\gamma}[\falg_3(q^{-1})]
X_{\alpha\gamma}[\balg_2(q)]
X_{\alpha\underline{\beta}}[\falg_1(q)]
\end{equation}
gives uniquely
\begin{equation}\label{Rfaf-1}
\Rop\;=\;\Rop_{\falg_1(q)\balg_2(q)\falg_3(q^{-1})}=
v^{\bosn_2}\rop_{\falg_1\balg_2\falg_3}^{} u^{\fern_1}
w^{-\fern_3}
\end{equation}
where
\begin{equation}\label{rfaf}
\begin{array}{ll}
\ds\rop_{\falg_1\balg_2\falg_3}=& \ds (-q)^{-\bosn_2}\biggl(
(1-\fern_1)k_2(1-\fern_3)+q^{-1}\fern_1(1-\fern_3) +
(1-\fern_1)\fern_3 + \fern_1k_2\fern_3\\
&\\
&\ds
+\frac{q^{-1}\fer_1^+\bos_2^{-}\fer_3^{-}-\fer_1^{-}\bos_2^+\fer_3^+}{1-q^2}\biggr)
\end{array}
\end{equation}
Operator (\ref{rfaf}) is the root of unity but it is not symmetric
with respect to the anti-involution.

\subsection{Configuration
$\underline{\alpha}\beta\underline{\gamma}$}

Relation
\begin{equation}
X_{\underline{\alpha}\beta}[\falg_1(q^{-1})]
X_{\underline{\alpha\gamma}}[\balg_2(q^{-1})]
X_{\beta\underline{\gamma}}[\falg_3(q)]\ \Rop =\Rop \
X_{\beta\underline{\gamma}}[\falg_3(q)]
X_{\underline{\alpha\gamma}}[\balg_2(q^{-1})]
X_{\underline{\alpha}\beta}[\falg_1(q^{-1})]
\end{equation}
produces
\begin{equation}\label{Rfaf-2}
\Rop\;=\;\Rop_{\falg_1(q^{-1})\balg_2(q^{-1})\falg_3(q)}(u,v,w) =
v^{-\bosn_2} \rop_{\falg_1\balg_2\falg_3} u^{-\fern_1} w^{\fern_3}
\end{equation}
where $\rop_{\falg_1\balg_2\falg_3}^{}$ is given by (\ref{rfaf}).

\subsection{Remarks}

All constant $\rop$-matrices are roots of unity since for
$\lambda=\mu=1$
\begin{equation}
X[\alg]=X[\alg]^{-1}
\end{equation}
for all $X$-matrixes (\ref{xbalg1}-\ref{xfalg2}). Also, matrices
$X[\balg(q)]$, $X[\balg(q^{-1})]$, $X[\falg(q)]$ at
$\lambda=\mu=1$ and matrix $X[\falg(q^{-1})]$ at
$\lambda=1,\mu=-1$ are symmetric with respect to the
anti-involution $\bos^{\pm}\to\bos^{\mp}$ and
$\fer^{\pm}\to\fer^{\mp}$ accompanied by the matrix transposition.
Recall, this anti-involution is the Hermitian conjugation for
$0<q<1$ for Fermi oscillators and Bose oscillator in
representation (\ref{fock}). Representation (\ref{anti-fock}) of
Bose oscillator admits $(\bos^{\pm})^\dagger=-\bos^{\mp}$. Thus,
the unitarity of intertwiners is an extra condition fixing a
proper choice of representations (\ref{fock}) or
(\ref{anti-fock}). For instance, $\rop_{\falg_1,\falg_2,\balg_3}$
and $\rop_{\balg_1,\falg_2,\falg_3}$ are unitary for
representation (\ref{fock}) for $\balg_1$ and $\balg_3$ while
$\rop_{\falg_1,\balg_2,\falg_3}$ is unitary for representation
(\ref{anti-fock}) of $\balg_2$. Matrix
$\rop_{\balg_1,\balg_2,\balg_3}$ is unitary if representation
(\ref{anti-fock}) is chosen in even number of components
$\balg_1,\balg_2,\balg_3$.

\section{Examples of tetrahedra}

We have four constant $\rop$-matrices, spectral parameters enter
as simple exponential fields:
\begin{equation}\label{anyR}
\Rop_{\mathcal{C}_1\mathcal{C}_2\mathcal{C}_3}(u,v,w)= v^{\Nop_2}
\rop_{\mathcal{C}_1\mathcal{C}_2\mathcal{C}_3} u^{\Nop_1}
w^{\Nop_3}\;,
\end{equation}
where
\begin{center}
\begin{tabular}[b]{|c|c|c|c|c|}
\hline & $\balg(q)$ & $\balg(q^{-1})$ & $\falg(q)$ &
$\falg(q^{-1})$
\\\hline $\Nop$ & $\bosn$ & $-\bosn$ & $\fern$ & $-\fern$
\\\hline
\end{tabular}
\end{center}
The only difference between $\mathcal{C}(q)$ and
$\mathcal{C}(q^{-1})$ is the sign of a field exponent.

Note, any matrix $\Rop_{\mathcal{C}_1\mathcal{C}_2\mathcal{C}_3}$
commutes with $\Nop_1+\Nop_2$ and $\Nop_2+\Nop_3$. Therefore, the
spectral parameters can be removed from any tetrahedron equation
and finally we have only eight constant tetrahedron equations
\begin{equation}
\rop_{\mathcal{C}_1\mathcal{C}_2\mathcal{C}_3}
\rop_{\mathcal{C}_1\mathcal{C}_4\mathcal{C}_5}
\rop_{\mathcal{C}_2\mathcal{C}_4\mathcal{C}_6}
\rop_{\mathcal{C}_3\mathcal{C}_5\mathcal{C}_6} =
\rop_{\mathcal{C}_3\mathcal{C}_5\mathcal{C}_6}
\rop_{\mathcal{C}_2\mathcal{C}_4\mathcal{C}_6}
\rop_{\mathcal{C}_1\mathcal{C}_4\mathcal{C}_5}
\rop_{\mathcal{C}_1\mathcal{C}_2\mathcal{C}_3}
\end{equation}
with
\begin{center}
\begin{tabular}[b]{|c|c|c|c|c|c|c|c|c|}
\hline Variant & $\mathcal{C}_1$ & $\mathcal{C}_2$ &
$\mathcal{C}_3$ &
$\mathcal{C}_4$ & $\mathcal{C}_5$ & $\mathcal{C}_6$ \\
\hline\hline 1 &
$\balg_1$ & $\balg_2$ & $\balg_3$ & $\balg_4$ & $\balg_5$ & $\balg_6$ \\
\hline\hline 2 &
$\balg_1$ & $\balg_2$ & $\balg_3$ & $\falg_4$ & $\falg_5$ & $\falg_6$ \\
\hline 3 &
$\balg_1$ & $\falg_2$ & $\falg_3$ & $\balg_4$ & $\balg_5$ & $\falg_6$ \\
\hline 4 &
$\falg_1$ & $\balg_2$ & $\falg_3$ & $\balg_4$ & $\falg_5$ & $\balg_6$ \\
\hline 5 &
$\falg_1$ & $\falg_2$ & $\balg_3$ & $\falg_4$ & $\balg_5$ & $\balg_6$ \\
\hline\hline 6 &
$\balg_1$ & $\falg_2$ & $\falg_3$ & $\falg_4$ & $\falg_5$ & $\balg_6$ \\
\hline 7 &
$\falg_1$ & $\balg_2$ & $\falg_3$ & $\falg_4$ & $\balg_5$ & $\falg_6$ \\
\hline 8 &
$\falg_1$ & $\falg_2$ & $\balg_3$ & $\balg_4$ & $\falg_5$ & $\falg_6$ \\
\hline
\end{tabular}
\end{center}

\subsection{Tetrahedron $\balg\balg\balg\falg\falg\falg$}

The quadrilateral configuration
\begin{equation}
X_{\alpha\beta}[\balg_1^{}] X_{\alpha\gamma}[\balg_2^{}]
X_{\beta\gamma}[\balg_3^{}]
X_{\alpha\underline{\delta}}[\falg_4^{}]
X_{\beta\underline{\delta}}[\falg_5^{}]
X_{\gamma\underline{\delta}}[\falg_6^{}]
\end{equation}
provides the tetrahedron equation
\begin{equation}\label{RLLL}
\rop_{\balg_1\balg_2\balg_3} \rop_{\balg_1\falg_4\falg_5}
\rop_{\balg_2\falg_4\falg_6} \rop_{\balg_3\falg_5\falg_6} =
\rop_{\balg_3\falg_5\falg_6} \rop_{\balg_2\falg_4\falg_6}
\rop_{\balg_1\falg_4\falg_5} \rop_{\balg_1\balg_2\balg_3}\;,
\end{equation}
number 2 from the table. Being written in matrix elements in
fermionic spaces, $|1\rangle \sim \fer^+|0\rangle$,
\begin{equation}
\rop_{\balg_1\falg_2\falg_3}\ id\otimes |j_2\rangle \otimes
|j_3\rangle \;=\; \sum_{i_1,i_2} |i_2\rangle\otimes |i_3\rangle
(-)^{p(i_1)p(i_2)} L_{i_1,i_2}^{j_1,j_2}[\balg_1]\;,
\end{equation}
where fermionic occupation numbers $i,j=0,1$ and the parity is
$p(i)=i$, equation (\ref{RLLL}) is equivalent to $RLLL$ relation
(4) from \cite{BS05}.

\subsection{Tetrahedron $\balg\falg\falg\falg\falg\balg$}

The quadrilateral configuration
\begin{equation}
X_{\underline{\alpha\beta}}[\balg_1^{}]
X_{\underline{\alpha}\gamma}[\falg_2^{}]
X_{\underline{\beta}\gamma}[\falg_3]
X_{\underline{\alpha}\delta}[\falg_4^{}]
X_{\underline{\beta}\delta}[\falg_5^{}]
X_{\gamma\delta}[\balg_6^{}]
\end{equation}
provides the constant tetrahedron equation
\begin{equation}\label{LLMM}
\rop_{\balg_1\falg_2\falg_3} \rop_{\balg_1\falg_4\falg_5}
\rop_{\falg_2\falg_4\balg_6} \rop_{\falg_3\falg_5\balg_6} =
\rop_{\falg_3\falg_5\balg_6} \rop_{\falg_2\falg_4\balg_6}
\rop_{\balg_1\falg_4\falg_5} \rop_{\balg_1\falg_2\falg_3}\;,
\end{equation}
number 6 from the table above. Fermionic lines in this equation
are not planar, hence sign parity factors are irremovable from a
matrix form of (\ref{LLMM}). However, the parity factors can be
absorbed by a re-definition of $\balg_6$, result is the
``strange'' $LLMM$ relation (34) from \cite{BS05}.

\section{Yang-Baxter equation and transfer matrices}

\subsection{$R$-matrices of Yang-Baxter equation}

An $R$-matrix of the Yang-Baxter equation can be defined by
\begin{equation}\label{Rybe}
\Ryb_{\alg_1,\alg_2}\;=\;\mathop{\textrm{Trace}}_{\mathcal{C}_{3}}
\left(\prod^\curvearrowright_{\ell}
\Rop_{\mathcal{C}_{1:\ell}\mathcal{C}_{2:\ell},\mathcal{C}_{3}}\right)\;,
\end{equation}
where semicolon in indices separates the orientation index
$j=1,2,3$ and the coordinate indices, the ordered product stands
for
\begin{equation}
\prod^\curvearrowright_{\ell} f_\ell\;=\;f_1f_2\cdots f_L\;,
\end{equation}
and the spaces of two-dimensional $R$-matrix are
\begin{equation}
\alg_1=\mathop{\textrm{\LARGE $\otimes$}}_{\ell=1}^L
\alg_{1:\ell}\;,\quad \alg_2=\mathop{\textrm{\LARGE
$\otimes$}}_{\ell=1}^L \alg_{2:\ell}\;.
\end{equation}
The sequence of tetrahedron equations
\begin{equation}
\left(\prod^\curvearrowright_{\ell}
\Rop_{\alg_{1:\ell}\alg_{2:\ell}\alg_3}
\Rop_{\alg_{1:\ell}\alg_{4:\ell}\alg_5}
\Rop_{\alg_{2:\ell}\alg_{4:\ell}\alg_6}\right)
\Rop_{\alg_{3}\alg_{5}\alg_6} = \Rop_{\alg_{3}\alg_{5}\alg_6}
\left(\prod^\curvearrowright_{\ell}
\Rop_{\alg_{2:\ell}\alg_{4:\ell}\alg_6}
\Rop_{\alg_{1:\ell}\alg_{4:\ell}\alg_5}
\Rop_{\alg_{1:\ell}\alg_{2:\ell}\alg_3}\right)
\end{equation}
provide the Yang-Baxter equation for (\ref{Rybe}),
\begin{equation}\label{YBE-gen}
\Ryb_{\alg_1\alg_2} \Ryb_{\alg_1\alg_4} \Ryb_{\alg_2\alg_4} \;=\;
\Ryb_{\alg_2\alg_4} \Ryb_{\alg_1\alg_4} \Ryb_{\alg_1\alg_2}\;.
\end{equation}
The ``third'' space of three-dimensional $R$-matrix is chosen as
the hidden space in (\ref{Rybe}), in general it can be ``first''
or ``second'' orientation spaces as well.

Locally, the spectral parameters enter (\ref{Rybe}) as
\begin{equation}
\Rop_{\mathcal{C}_{1:\ell}\mathcal{C}_{2:\ell},\mathcal{C}_{3}}\;=\;
v_\ell^{\Nop_{2:\ell}}
\rop_{\mathcal{C}_{1:\ell}\mathcal{C}_{2:\ell},\mathcal{C}_{3}}
u_\ell^{\Nop_{1:\ell}} w_\ell^{\Nop_{3}}\;,
\end{equation}
see (\ref{anyR}), where
\begin{equation}
u_\ell=\frac{\lambda_3}{\lambda_{2:\ell}}\;,\quad
v_\ell=\lambda_{1:\ell}\mu_3\;,\quad
w_\ell=\frac{\mu_{1:\ell}}{\mu_{2:\ell}}\;.
\end{equation}
Since $\Nop_1+\Nop_2$ and $\Nop_2+\Nop_3$ commute with
$\rop_{\alg_1\alg_2\alg_3}$, the expression for (\ref{Rybe}) can
be rewritten identically as
\begin{equation}\label{105}
\Ryb_{\alg_1\alg_2} = U_1^{-1}U_2^{-1}\ \left(\prod_\ell
v_\ell^{\Nop_{2:\ell}} \right)\ \Ryb_{\alg_1\alg_2}(\mu_1/\mu_2)\
\left(\prod_\ell u_\ell^{\Nop_{1:\ell}}\right)\ U_1^{}U_2^{}\;,
\end{equation}
where the gauge factors are
\begin{equation}
U_1\;=\; \prod_{\ell=1}^{L}
\mu_{1:\ell}^{\sum_{\ell'=0}^{\ell}\Nop_{1:\ell'}}\;,\quad
U_2\;=\; \prod_{\ell=1}^{L}
\mu_{2:\ell}^{\sum_{\ell'=0}^{\ell}\Nop_{2:\ell'}}\;,
\end{equation}
the simplified $R$-matrix in the right hand side of (\ref{105})
is\footnote{Due to the factor $w^{\Nop_3}$, there are no necessity
to use super-trace for the case of $\alg_3=\falg_3$.}
\begin{equation}\label{Rybe-w}
\Ryb_{\alg_1,\alg_2}(w)\;=\;\mathop{\textrm{Trace}}_{\mathcal{C}_{3}}
\left(w^{\Nop_3}\prod^\curvearrowright_{\ell}
\rop_{\mathcal{C}_{1:\ell}\mathcal{C}_{2:\ell},\mathcal{C}_{3}}\right)\;,
\end{equation}
and its single spectral parameter is given by
\begin{equation}
\mu_1/\mu_2 \;=\;\prod_\ell w_\ell\;,\quad
\mu_1=\prod_\ell\mu_{1:\ell}\;,\quad
\mu_2=\prod_\ell\mu_{2:\ell}\;.
\end{equation}
$R$-matrix (\ref{Rybe}) commutes with
$\Nop_{1:\ell}+\Nop_{2:\ell}$ for any $\ell$ and with
\begin{equation}\label{centers}
\Nop_{1:*}=\sum_\ell \Nop_{1:\ell}\quad\textrm{and}\quad
\Nop_{2:*}=\sum_\ell \Nop_{2:\ell}\;,
\end{equation}
what corresponds to arbitrariness of parameters $\lambda_3,\mu_3$
of hidden space. Cancelling then all gauges and fields in the
general Yang-Baxter equation (\ref{YBE-gen}), we come to the
standard Yang-Baxter equation with multiplicative spectral
parameter for simplified $R$-matrix (\ref{Rybe-w})
\begin{equation}
\Ryb_{\alg_1\alg_2}(\mu_1/\mu_2) \Ryb_{\alg_1\alg_4}(\mu_1/\mu_4)
\Ryb_{\alg_2\alg_4}(\mu_2/\mu_4) =
\Ryb_{\alg_2\alg_4}(\mu_2/\mu_4) \Ryb_{\alg_1\alg_4}(\mu_1/\mu_4)
\Ryb_{\alg_1\alg_2}(\mu_1/\mu_2)\;.
\end{equation}

\subsection{Layer-to-layer transfer matrices}

For a given set of quantum spaces $\alg_{1:\ell,m}$ and for any
suitable sequences of auxiliary spaces $\mathcal{C}_{2:\ell}$ and
$\mathcal{C}_{3:m}$, the layer-to-layer transfer matrix is
\begin{equation}\label{tm}
\mathbf{T}\;=\;\mathop{\textrm{Trace}}_{\mathcal{C}_{2:\ell},\mathcal{C}_{3:m}}
\biggl( \prod^\curvearrowright_{\ell}\prod^\curvearrowright_m
\Rop_{\mathcal{C}_{1:\ell,m}\mathcal{C}_{2:\ell},\mathcal{C}_{3:m}}
\biggr)\;,
\end{equation}
where the ordered products are defined by
\begin{equation}
\prod^\curvearrowright_{\ell} f_\ell\;=\;f_1f_2\cdots f_L\;,\quad
\prod^\curvearrowright_{m} g_m\;=\;g_1g_2\cdots g_M\;.
\end{equation}
Transfer matrix (\ref{tm}) can be understood as a two-dimensional
transfer matrix for a length $M$ chain of $R$-matrices
(\ref{Rybe}) with hidden ``third'' spaces, or as a two-dimensional
transfer matrix for a length $L$ chain of $R$-matrices with hidden
``second'' spaces. The tetrahedron equation provides the
commutativity of any two transfer matrices with identical sets of
$\alg_{1:\ell,m}$. Transfer matrices can differ by spectral
parameters in auxiliary spaces and by a choice of statistics of
auxiliary spaces.

Locally, the spectral parameters enter (\ref{tm}) as
\begin{equation}
\Rop_{\mathcal{C}_{1:\ell,m}\mathcal{C}_{2:\ell},\mathcal{C}_{3:m}}\;=\;
(\lambda_{1:\ell,m}\mu_{3:m})^{\Nop_{2:\ell}}
\rop_{\mathcal{C}_{1:\ell,m}\mathcal{C}_{2:\ell},\mathcal{C}_{3:m}}
\left(\frac{\lambda_{3:m}}{\lambda_{2:\ell}}\right)^{\Nop_{1:\ell,m}}
\left(\frac{\mu_{1:\ell,m}}{\mu_{2:\ell}}\right)^{\Nop_{3:m}}\;,
\end{equation}
see (\ref{anyR}). Since $\Nop_1+\Nop_2$ and $\Nop_2+\Nop_3$
commute with $\rop_{\alg_1\alg_2\alg_3}$, the spectral parameters
in auxiliary spaces can be pushed to boundary, transfer matrix
(\ref{tm}) can be rewritten as
\begin{equation}\label{tm2}
\mathbf{T}\;=\;\mathop{\textrm{Trace}}_{\mathcal{C}_{2:\ell},\mathcal{C}_{3:m}}
\biggl( \prod_\ell v_\ell^{\Nop_{2:\ell}} \; \prod_m
w_m^{\Nop_{3:m}}\;
\prod^\curvearrowright_{\ell}\prod^\curvearrowright_m
\rop_{\mathcal{C}_{1:\ell,m}\mathcal{C}_{2:\ell},\mathcal{C}_{3:m}}
\biggr) \prod_{\ell,m}
\left(\frac{\lambda_{3:m}}{\lambda_{2:\ell}}\right)^{\Nop_{1:\ell,m}}\;,
\end{equation}
where
\begin{equation}
v_\ell\;=\;\prod_m \lambda_{1:\ell,m}\mu_{3:m}\;,\quad
w_m=\prod_\ell\frac{\mu_{1:\ell,m}}{\mu_{2:\ell}}
\end{equation}
This most general expression corresponds to inhomogeneous spectral
parameters $v_\ell\neq v_{\ell'}$, $w_m\neq w_{m'}$. The most
right external factor in (\ref{tm2}) depends only on auxiliary
spectral parameters, hence
\begin{equation}
\Nop_{\ell*}=\sum_m \Nop_{1:\ell,m}\quad \textrm{and}\quad
\Nop_{*m}=\sum_\ell \Nop_{1:\ell,m}
\end{equation}
commute with transfer matrix and therefore this factor is
inessential for spectral problem.

The choice $\lambda_{1:\ell,m}=\mu_{1:\ell,m}$ gives the
homogeneous transfer matrix,
\begin{equation}\label{tm3}
\mathbf{T}(v,w)\;=\;\mathop{\textrm{Trace}}_{\mathcal{C}_{2},\mathcal{C}_{3}}
\biggl( v^{\Nop_{2}}  w^{\Nop_{3}}\;
\prod^\curvearrowright_{\ell}\prod^\curvearrowright_m
\rop_{\mathcal{C}_{1:\ell,m}\mathcal{C}_{2:\ell},\mathcal{C}_{3:m}}
\biggr)\;,
\end{equation}
where for shortness
\begin{equation}
\Nop_2=\sum_\ell \Nop_{2:\ell}\quad\textrm{and}\quad
\Nop_3=\sum_m\Nop_{3:m}\;.
\end{equation}
The tetrahedron equations and related effective Yang-Baxter
equations provide the commutativity of transfer matrices,
\begin{equation}
\biggl[\mathbf{T}(v,w),\mathbf{T}(v',w')\biggr]\;=\;0\;.
\end{equation}

\section{Classification of $R$-matrices in terms quantum groups}

\subsection{General case}
Consider effective two-dimensional $R$-matrix (\ref{Rybe-w}) with
$\alg_3=\balg_3(q)$. Let all Bose $q$-oscillators are at unitary
representation (\ref{fock}). According to the patterns
(\ref{Raaa-1}) and (\ref{Rffa-1}), possible choice of
$\alg_{1:\ell}$ and $\alg_{2:\ell}$ are
\begin{equation}
\alg_{1:\ell}\otimes \alg_{2:\ell} \;=\;\left[\begin{array}{l}
\ds \balg_{1:\ell}(q)\otimes \balg_{2:\ell}(q)\\
\ds \textrm{or} \\
\ds \falg_{1:\ell}(q^{-1})\otimes \falg_{2:\ell}(q^{-1})
\end{array}\right.
\end{equation}
Thus, we can define the following space of the Yang-Baxter
equation:
\begin{equation}\label{Qspace}
Q_j\;=\;\mathop{\textrm{\LARGE $\otimes$}}_{\ell=1}^L
\left[\balg_{j:\ell}(q)\;\;\textrm{or}\;\;\falg_{j:\ell}(q^{-1})\right]\;,\quad
\# \balg_{j:\ell}=L_1\;,\quad \#\falg_{j:\ell}=L_2\;.
\end{equation}
Alternative space can be defined by
\begin{equation}\label{Aspace}
A_j\;=\;\mathop{\textrm{\LARGE $\otimes$}}_{\ell=1}^L
\left[\falg_{j:\ell}(q)\;\;\textrm{or}\;\;\balg_{j:\ell}(q^{-1})\quad
\textrm{respectively} \right]\;.
\end{equation}
Any of choices $\alg_j=Q_j$ or $A_j$ in the Yang-Baxter equation
(\ref{YBE-gen}) is valid. Matrix $\Rop_{Q_1,Q_2}$ has the hidden
space $\balg_3(q)$, matrix $\Rop_{Q_1,A_2}$ has the hidden space
$\falg_3(q)$, matrix $R_{A_1,A_2}$ has the hidden space
$\balg_3(q^{-1})$, matrix $R_{A_1,Q_2}$ has hidden space
$\falg_3(q^{-1})$.

Our statement is the following: For given $L_1$ and $L_2$, all the
$R$-matrices reproduce the $R$-matrices and $L$-operators for
quantum super-algebra
$\mathscr{U}_q(\widehat{\textrm{gl}}(L_1|L_2))$. Spaces $Q$ and
$A$ are two types of \emph{reducible} evaluation representations
of $\mathscr{U}_q(\widehat{\textrm{gl}}(L_1|L_2))$
\cite{Kac:1977L}.

If $L_2=0$, these $R$-matrices correspond to
$\mathscr{U}_q(\widehat{sl}_L)$. It is shown in \cite{BS05}, the
space $Q_j$ is the direct infinite sum of all symmetric tensor
representations of $sl_L$; the space $A_j$ is the direct sum of
all antisymmetric tensor (fundamental) representations of $sl_L$.
Note, in a case of a mixture of representations (\ref{fock}) and
(\ref{anti-fock}), infinite dimensional evaluation representations
of $\mathscr{U}_q(\widehat{sl}_L)$ appear; we do not discuss this
possibility in details here.

Remarks:
\begin{itemize}
\item For given $L_1$ and $L_2$, there are different choices of
particular ordering of $\balg$ and $\falg$. This corresponds to
different choices of Cartan matrix and Dynkin diagram for
$\textrm{gl}(L_1|L_2)$. All such choices are equivalent due to the
tetrahedron equation (providing Z-invariance of three-dimensional
lattice). \item Transfer matrix (\ref{tm}) has the structure of
simple square lattice in auxiliary orientation spaces ``2'' and
``3''. More complicated structure of auxiliary configuration
provides more reach class of evaluation representations, e.g.
multicomponent Bose/Fermi gases.
\end{itemize}
The case of massless free fermions $L_1=L_2=1$ is too primitive.
Below we argue our statement for more illustrative case $L_1=2$,
$L_2=1$.

\subsection{$R$-matrices for $\mathscr{U}_q(\widehat{\textrm{gl}}(2|1))$}

For $L_1=2$ and $L_2=1$ the ``quantum'' and ``auxiliary'' spaces
(\ref{Qspace},\ref{Aspace}) are respectively
\begin{equation}
Q_j\;=\;\balg_{j:1}(q)\otimes \balg_{j:2}(q)\otimes
\falg_{j:3}(q^{-1})\;,\quad A_j\;=\;\falg_{j:1}(q)\otimes
\falg_{j:2}(q)\otimes \balg_{j:3}(q^{-1})\;.
\end{equation}
These spaces have the following invariants (\ref{centers}):
\begin{equation}
\Nop_{Q_j}=\bosn_{j:1}+\bosn_{j:2}-\fern_{j:3}\;,\quad
\Nop_{A_j}=\fern_{j:1}+\fern_{j:2}-\bosn_{j:3}\;.
\end{equation}
Both $Q$ and $A$ are infinite dimensional spaces. Spectrum of
$\Nop_Q$ is $-1,0,1,2,3,\dots$. Let $Q(N)$ be a subspace of $Q$
with $\Nop_Q=N$. Elementary combinatorics gives
\begin{equation}
\dim Q(N)=2N+3\;.
\end{equation}
We identify $Q(-1)$ -- scalar representation, $Q(0)$ -- vector
representation, $Q(N)$ with $N\geq 1$ -- highest atypical
representations of $gl(2|1)$.

In its turn, spectrum of $\Nop_A$ is $2,1,0,-1,\dots$. Let again
$A(N)$ be a subspace of $A$ with $\Nop_A=N$. Then
\begin{equation}
\dim A(2)=0\;,\quad \dim A(1)=3\;,\quad \dim
A(N)=4\;\;\textrm{for}\;\; N\leq 0\;.
\end{equation}
We identify $A(2)$ -- scalar representation, $A(1)$ -- vector
representation, $A(N)$ with $N\leq 0$ -- typical representation of
$gl(2|1)$.

Below we consider two $R$-matrices,
\begin{equation}
\Ryb_{A_1,Q_2}(w)\;=\;(-q)^{\Nop_{Q_2}}\
\mathop{\textrm{Trace}}_{\falg_3} \left( w^{\fern_3}
\rop_{\falg_{1:1}\balg_{2:1}\falg_3}
\rop_{\falg_{1:2}\balg_{2:2}\falg_3}
\rop_{\balg_{1:3}\falg_{2:3}\falg_3} \right)
\end{equation}
and
\begin{equation}
\Ryb_{A_1,A_2}(w)\;=\;\mathop{\textrm{Trace}}_{\balg_3} \left(
w^{\bosn_3} \rop_{\falg_{1:1}\falg_{2:1}\balg_3}
\rop_{\falg_{1:2}\falg_{2:2}\balg_3}
\rop_{\balg_{1:3}\balg_{2:3}\balg_3} \right)
\end{equation}
Explicit expression for $\Ryb_{A_1,Q_2}(w)$ is
\begin{equation}
\begin{array}{l}
\ds \Ryb_{A_1,Q_2}(w) =\\
\\
\ds  [(1-\fern_{11})\kop_{21}+q^{-1}\fern_{11}]
[(1-\fern_{12})\kop_{22}+q^{-1}\fern_{12}]
[(1-\fern_{23})+\kop_{13}\fern_{23}]\\
\\
\ds + w [(1-\fern_{11})+\fern_{11}\kop_{21}]
[(1-\fern_{12})+\fern_{12}\kop_{22}]
[\kop_{13}(1-\fern_{23})+q^{-1}\fern_{23}]\\
\\
\ds
+\frac{q^{-1}}{1-q^2}\fer_{11}^+\fer_{12}^{-}\bos_{21}^{}\bos_{22}^+
[(1-\fern_{23})+\kop_{13}\fern_{23}] +
w\frac{q^{-1}}{1-q^2}\fer_{11}^{-}\fer_{12}^+\bos_{21}^+\bos_{22}^{-}
[\kop_{13}(1-\fern_{23})+q^{-1}\fern_{23}]\\
\\
\ds +\frac{q^{-1}}{1-q^2}\fer_{11}^+\bos_{21}^{-}
[(1-\fern_{12})+\fern_{12}\kop_{22}] \bos_{13}^+\fer_{23}^{-} -
w\frac{q^{-1}}{1-q^2} \fer_{11}^{-}\bos_{21}^+
[(1-\fern_{12})\kop_{22}+q^{-1}\fern_{12}]
\bos_{13}^{-}\fer_{23}^+\\
\\
\ds
+\frac{q^{-1}}{1-q^2}[(1-\fern_{11})\kop_{21}+q^{-1}\fern_{11}]
\fer_{12}^+\bos_{22}^{-}\bos_{13}^+\fer_{23}^{-} - w
\frac{q^{-1}}{1-q^2} [(1-\fern_{11})+\fern_{11}\kop_{21}]
\fer_{12}^{-}\bos_{22}^+\bos_{13}^{-}\fer_{23}^+
\end{array}
\end{equation}
In what follows, it is more convenient to change notations to
tensor product form:
\begin{equation}
\alg_{1;j}\alg_{2:k} = \alg_j\otimes \alg_k
\end{equation}

Consider now the following basis of $A_1=A\otimes 1$ with
$\Nop_{A}=-n$ (recall, we consider now the unitary representation
(\ref{fock}) of Bose oscillator, $|0\rangle$ is the total even
Fock vacuum annihilated by all $\fer^-$ and $\bos^-$ operators):
\begin{equation}
\begin{array}{l}
\ds  |e_0\rangle \;=\;\frac{\bos_3^{+
n}}{\sqrt{(q^2;q^2)_n}}|0\rangle = |0,0,n\rangle\;,\\
\\
\ds |e_1\rangle = \frac{\fer_1^+}{\sqrt{1-q^2}}\frac{\bos_3^{+
(n+1)}}{\sqrt{(q^2;q^2)_{n+1}}}|0\rangle =
|1,0,n+1\rangle\;,\\
\\
\ds |e_2\rangle = \frac{\fer_2^+}{\sqrt{1-q^2}}\frac{\bos_3^{+
(n+1)}}{\sqrt{(q^2;q^2)_{n+1}}}|0\rangle=|0,1,n+1\rangle\;,\\
\\
\ds |e_3\rangle = \frac{\fer_1^+\fer_2^+}{1-q^2}\frac{\bos_3^{+
(n+2)}}{\sqrt{(q^2;q^2)_{n+2}}}|0\rangle=|1,1,n+2\rangle\;.
\end{array}
\end{equation}
Parity of states are $p(e_0)=p(e_3)=0$ and $p(e_1)=p(e_2)=1$.
Matrix units $E_{jk}$ can be introduced by
\begin{equation}
E_{jk}|e_k\rangle = |e_j\rangle
\end{equation}
Using definition of matrix units in $A\otimes 1$ space, we rewrite
operator $\Ryb_{A\otimes Q}$ as {\small
\begin{equation}
\begin{array}{l}
\ds\Ryb_{A\otimes Q}\;=\; E_{00} \otimes
\left(q^{\bosn_1+\bosn_2+n\fern_3}+wq^{n-(1+n)\fern_3}\right) +
E_{11}\otimes \left(q^{\bosn_2+(n+1)\fern_3-1} + w
q^{n+1+\bosn_1-(n+2)\fern_3}\right)\\
\\
+E_{22}\otimes \left(q^{\bosn_1+(n+1)\fern_3-1} + w
q^{n+1+\bosn_2-(n+2)\fern_3}\right) + E_{33}\otimes
\left(q^{(n+2)\fern_3-2} + w
q^{n+2+\bosn_1+\bosn_2-(n+3)\fern_3}\right)\\
\\
\ds + E_{12}\otimes \bos_1^{}\bos_2^+ q^{(n+1)\fern_3-1} - w
E_{21}\otimes \bos_1^+\bos_2^{} q^{n-(n+2)\fern_3}\\
\\
\ds +q^{-1}\sqrt{\frac{1-q^{2(n+1)}}{1-q^2}} E_{10}\otimes
\bos_1^{}\fer_3^{} - q^{-1}\sqrt{\frac{1-q^{2(n+2)}}{1-q^2}}
E_{32}\otimes \bos_1q^{\bosn_2}\fer_3\\
\\
\ds +q^{-1} w \sqrt{\frac{1-q^{2(n+1)}}{1-q^2}} E_{01}\otimes
\bos_1^+ q^{\bosn_2}\fer_3^+ - q^{-2} w
\sqrt{\frac{1-q^{2(n+2)}}{1-q^2}} E_{23}\otimes
\bos_1^+\fer_3^+\\
\\
\ds + q^{-1}\sqrt{\frac{1-q^{2(n+1)}}{1-q^2}} E_{20}\otimes
q^{\bosn_1}\bos_2\fer_3 + q^{-2}
\sqrt{\frac{1-q^{2(n+2)}}{1-q^2}} E_{31}\otimes \bos_2\fer_3\\
\\
\ds + q^{-1} w \sqrt{\frac{1-q^{2(n+1)}}{1-q^2}} E_{02}\otimes
\bos_2^+ \fer_3^+ + q^{-1} w \sqrt{\frac{1-q^{2(n+2)}}{1-q^2}}
E_{13}\otimes q^{\bosn_1}\bos_2^+ \fer_3^+
\end{array}
\end{equation}}
When $n=-1$, $E_{00}$ component factors out and one has
three-dimensional representation in $A$-space:
\begin{equation}
L_{A\otimes Q}(u)\;=\;q R_{A\otimes
Q}(w=-uq^{\bosn_1+\bosn_2+1-\fern_3})\;=\;\sum_{j,k=1}^3
E_{jk}\otimes A_{kj}(u)\;,
\end{equation}
where
\begin{equation}
A_{11}=q^{\bosn_2}-uq^{-\bosn_2}\;,\quad
A_{22}=q^{\bosn_1}-uq^{-\bosn_1}\;,\quad
A_{33}=q^{\fern_3-1}-uq^{1-\fern_3}\;,
\end{equation}
and
\begin{equation}
\begin{array}{lll}
\ds A_{12}=uq^{-(\bosn_1+\bosn_2+1)}\bos_1^+\bos_2^{}\;,& \ds
A_{31}=-uq^{-\bosn_2}\bos_2^+ \fer_3^+\;, & \ds
A_{32}=uq^{-(\bosn_1+\bosn_2+1)}\bos_1^+\fer_3^+\;,\\
&&\\
\ds A_{21}=\bos_1^{}\bos_2^+\;,  & \ds
A_{13}=q^{-1}\bos_2^{}\fer_3^{}\;, & \ds A_{23}=-
q^{\bosn_2}\bos_1^{}\fer_3^{}\;.
\end{array}
\end{equation}
This is definitely the $L$-operator for
$\mathscr{U}_q(\widehat{\textrm{gl}}(2|1))$ with vector
representation in auxiliary space $A$ and oscillator evaluation
representation \cite{Kulish:1990S} in quantum space $Q$. A few
exchange relations for $A_{ij}$,
\begin{equation}
\begin{array}{l}
\ds
[A_{12},A_{21}]=u(q^{-1}-q)(q^{\bosn_2-\bosn_1}-q^{\bosn_1-\bosn_2})\;,\\
\\
\ds
[A_{31},A_{13}]_+=u(q^{-1}-q)(q^{\bosn_2+1-\fern_3}-q^{-\bosn_2-1+\fern_3})\;,\\
\\
\ds
[A_{32},A_{23}]_+=u(q^{-1}-q)(q^{\bosn_1+1-\fern_3}-q^{-\bosn_1-1+\fern_3})\;,
\end{array}
\end{equation}
fix the Cartan elements of $\textrm{gl}(2|1)$
\begin{equation}
h_1=\bosn_2+1-\fern_3\;,\quad h_2=\bosn_1-\bosn_2\;,\quad
h_3=h_1+h_2\;.
\end{equation}
In its turn, all sixteen matrix elements of operator
$\Ryb_{A_1,A_2}(w)$
\begin{equation}\label{Rtyptyp}
\Ryb_{A_1,A_2}(w) |e_{j_1}\rangle \otimes |e_{j_2}\rangle \;=\;
\sum_{k_1,k_2} |e_{k_1}\rangle \otimes |e_{k_2}\rangle\
(-)^{p(k_1)p(k_2)}\ R_{k_1,k_2}^{j_1,j_2}(w)
\end{equation}
can be calculated explicitly with the help of generating functions
(\ref{tr3}) {\small
\begin{equation}
\begin{array}{l}
\ds \langle n_1+k,n_2|\mathop{\textrm{Tr}}_{\balg_3}\left(
v^{\bosn_3} \bos_3^k
\rop_{\balg_1\balg_2\balg_3}\right)|n_1,n_2+k\rangle = v^{n_2} \;
\sqrt{\frac{(q^2;q^2)_{n_1+k,n_2+k}}{(q^2;q^2)_{n_1,n_2}}}\;
\frac{(q^{n_1-n_2+2}v^{-1};q^2)_{n_2}}{(q^{n_1-n_2}v;q^2)_{n_2+k+1}}
\;,\\
\\
\ds \langle n_1,n_2+k|\mathop{\textrm{Tr}}_{\balg_3}\left(
v^{\bosn_3} \bos_3^{+ k}
\rop_{\balg_1\balg_2\balg_3}\right)|n_1+k,n_2\rangle = v^{n_2+k}
\; \sqrt{\frac{(q^2;q^2)_{n_1+k,n_2+k}}{(q^2;q^2)_{n_1,n_2}}}\;
\frac{(q^{n_1-n_2+2}v^{-1};q^2)_{n_2}}{(q^{n_1-n_2}v;q^2)_{n_2+k+1}}
\end{array}
\end{equation}}
Parameters $n_1=-\Nop_{A_1}$ and $n_2=-\Nop_{A_2}$ are the
additional spectral parameters for $\Ryb_{A_1,A_2}$ with no
difference property. Matrix (\ref{Rtyptyp}) coincides with that
from \cite{Bracken:1994I,Gould:1994II}.

\section{Conclusion}

In the algebraic approach, a two-dimensional quantum integrable
model is defined by a quantum group and by its evaluation
representation. Contrary to the two-dimensional case, such choice
for three-dimensional models is rather limited, instead of a rich
representation theory one has locally just the choice of
statistics: Bose or Fermi. Note however, three-dimensional
$R$-matrices intertwine even number of fermions, there is no
three-fermions intertwiners \cite{Bazhanov:1981zm,Ambjorn} in our
scheme. The transfer matrix in $3d$ is the layer-to-layer transfer
matrix, it has not a structure of a one-dimensional quantum chain
but the structure of a two-dimensional quantum lattice. It is
shown in this paper, the simple square quantum lattice reproduces
a collection of effective two-dimensional models for at least
quantum super-algebras of $\widehat{\textrm{gl}}$ type and certain
set of their representations. More complicated quantum lattices
\cite{Sergeev:PN} produces more complicated evaluation
representations of quantum groups, for instance the lattice Bose
and Fermi gases, their multi-component generalizations, etc., as
well as $q$-deformed Toda chain \cite{Kharchev:2002} and quantum
Liouville theory \cite{FKV:2001}. Moreover, there are specific
quantum lattices with twisted boundary that have no quantum group
interpretation at all \cite{Sergeev:2006}.

Classification theorems in this paper are essentially based on the
ultra-locality test. A three-dimensional analogue of fusion gives
a simple example when this test breaks down: the matrix
\begin{equation}
X_{\boldsymbol{\alpha\beta}}\;=\;
X_{\alpha_1\beta_1}[\balg_{11}]X_{\alpha_1\beta_2}[\balg_{12}]
X_{\alpha_2\beta_1}[\balg_{21}]X_{\alpha_2\beta_2}[\balg_{22}]
\end{equation}
is the block matrix, its elements $a,b,c,d$ are two-by-two blocks
with matrix elements from
$\balg_{11}\otimes\balg_{12}\otimes\balg_{21}\otimes\balg_{22}$.
The corresponding intertwiner of equation (\ref{xxxr}) is a
product of eight elementary intertwiners. Ultra-locality test does
not work for block matrices $a_j,b_j,c_j,d_j$ and thus the
classification scheme is essentially enlarged. A classification
method for the case when ultra-locality test is not applicable or
when involved algebras are not ultra-local is an open problem.

It worth noting here, the ``edge-type'' linear problem of Fig.
\ref{fig-lp} considered here is not only possible one; there is a
distinct quantum ``face-type'' linear problem providing the local
Weyl algebra of observables \cite{Sergeev:1999jpa,Sergeev:PN}; a
classification scheme for mixed quantum auxiliary linear problems
is not known either.

An inclusion of $B,C,D$ series into the tetrahedral scheme is not
yet known. However, there is no doubt that this is possible. For
instance, the spinor representations of rotation groups have
dimension $2^n$ what is an evident criterion of the hidden third
dimension.

\noindent\textbf{Acknowledgements.} I would like to thank Vladimir
Bazhanov,  Vladimir Mangazeev, Rinat Kashaev, Alexander Molev and
Jan de Gier for fruitful discussions. I am also grateful to Mary
Hewett, Judith Ascione and Peter Vassiliou from University of
Canberra for their friendly support.


\begin{thebibliography}{10}

\bibitem{BS05}
Bazhanov, V.~V. and Sergeev, S.~M.
\newblock Zamolodchikov's tetrahedron equation and hidden structure of quantum
  groups.
\newblock J. Phys. A {\bf 39} (2006) 3295--3310.

\bibitem{MR0457963}
Zaharov, V.~E. and Manakov, S.~V.
\newblock Generalization of the method of the inverse scattering problem.
\newblock Teoret. Mat. Fiz. {\bf 27} (1976) 283--287.

\bibitem{Melbourne}
Sergeev, S.
\newblock Quantization of three-wave equations.
\newblock J. Phys. A: Math. Theor. {\bf 40} (2007) 12709--12724.

\bibitem{circular}
Bazhanov, V.~V., Mangazeev, V.~V., and Sergeev, S.~M.
\newblock Quantum geometry of 3-dimensional lattices.
\newblock arXiv:0801.0129, 2008.

\bibitem{BogdanovKonopelchenko}
Bogdanov, L.~V. and Konopelchenko, B.~G.
\newblock Lattice and {$q$}-difference {D}arboux-{Z}akharov-{M}anakov systems
  via {$\overline\partial$}-dressing method.
\newblock J. Phys. A {\bf 28} (1995) L173--L178.

\bibitem{DoliwaSantini}
Doliwa, A. and Santini, P.~M.
\newblock Multidimensional quadrilateral lattices are integrable.
\newblock Phys. Lett. A {\bf 233} (1997) 365--372.

\bibitem{Korepanov:1995}
Korepanov, I.~G.
\newblock Algebraic integrable dynamical systems, $2+1$ dimensional models on
  wholly discrete space-time, and inhomogeneous models on 2-dimensional
  statistical physics.
\newblock arXiv:solv-int/9506003, 1995.

\bibitem{KashaevKorepanovSergeev}
Kashaev, R.~M., Korepanov, I.~G., and Sergeev, S.~M.
\newblock The functional tetrahedron equation.
\newblock Teoret. Mat. Fiz. {\bf 117} (1998) 370--384.

\bibitem{Sergeev:2005d}
Sergeev, S.
\newblock Quantum curve in {$q$}-oscillator model.
\newblock Int. J. Math. Math. Sci.  (2006) Art. ID 92064, 31.

\bibitem{GR}
Gasper, G. and Rahman, M.
\newblock {\em Basic Hyper-geometric Series}.
\newblock Cambridge University Press, Cambridge, 1990.

\bibitem{Shiroishi:1998F}
Umeno, Y., Shiroishi, M., and Wadati, M.
\newblock Fermionic {$R$}-operator for the fermion chain model.
\newblock J. Phys. Soc. Japan {\bf 67} (1998) 1930--1935.

\bibitem{Kac:1977L}
Kac, V.
\newblock Representations of classical {L}ie superalgebras.
\newblock In {\em Differential geometrical methods in mathematical physics, II
  (Proc. Conf., Univ. Bonn, Bonn, 1977)}, volume 676 of {\em Lecture Notes in
  Math.}, pages 597--626. Springer, Berlin, 1978.

\bibitem{Kulish:1990S}
Chaichian, M. and Kulish, P.
\newblock Quantum {L}ie superalgebras and {$q$}-oscillators.
\newblock Phys. Lett. B {\bf 234} (1990) 72--80.

\bibitem{Bracken:1994I}
Bracken, A.~J., Gould, M.~D., Zhang, Y.~Z., and Delius, G.~W.
\newblock Solutions of the quantum {Y}ang-{B}axter equation with extra
  non-additive parameters.
\newblock J. Phys. A {\bf 27} (1994) 6551--6561.

\bibitem{Gould:1994II}
Delius, G.~W., Gould, M.~D., Links, J.~R., and Zhang, Y.-Z.
\newblock Solutions of the {Y}ang-{B}axter equation with extra non-additive
  parameters. {II}. {$U\sb q({\rm gl}(m\vert n))$}.
\newblock J. Phys. A {\bf 28} (1995) 6203--6210.

\bibitem{Bazhanov:1981zm}
Bazhanov, V.~V. and Stroganov, Y.~G.
\newblock On commutativity conditions for transfer matrices on multidimensional
  lattice.
\newblock Theor. Math. Phys. {\bf 52} (1982) 685--691.

\bibitem{Ambjorn}
Ambjorn, J., Khachatryan, S., and Sedrakyan, A.
\newblock Simplified tetrahedron equations: Fermionic realization.
\newblock Nucl. Phys. {\bf B734} (2006) 287--303.

\bibitem{Sergeev:PN}
Sergeev, S.
\newblock {Q}uantum integrable models in discrete 2+1 dimensional space-time:
  auxiliary linear problem on a lattice, zero curvature representation,
  isospectral deformation of the {Z}amolodchikov-{B}azhanov-{B}axter model.
\newblock Particles and Nuclei {\bf 35} (2004) 1051--1111.

\bibitem{Kharchev:2002}
Kharchev, S., Lebedev, D., and Semenov-Tian-Shansky, M.
\newblock Unitary representations of $U_{q}(\mathfrak{sl}(2,\mathbb{R}))$, the
  modular double, and the multiparticle q-deformed {T}oda chains.
\newblock Commun. Math. Phys. {\bf 225} (2002) 573--609.

\bibitem{FKV:2001}
Faddeev, L.~D., Kashaev, R.~M., and Volkov, A.~Y.
\newblock Strongly coupled quantum discrete Liouville theory. I. Algebraic
  approach and duality.
\newblock Commun. Math. Phys. {\bf 219} (2001) 199--219.

\bibitem{Sergeev:2006}
Sergeev, S.
\newblock Ansatz of {H}ans {B}ethe for a two-dimensional lattice {B}ose gas.
\newblock J. Phys. A {\bf 39} (2006) 3035--3045.

\bibitem{Sergeev:1999jpa}
Sergeev, S.~M.
\newblock {Q}uantum {$2+1$} evolution model.
\newblock J. Phys. A: Math. Gen. {\bf 32} (1999) 5693--5714.

\end{thebibliography}
\end{document}